\documentclass{aastex631}

\def\ltsima{$\; \buildrel < \over \sim \;$}
\def\simlt{\lower.5ex\hbox{\ltsima}} 
\def\gtsima{$\; \buildrel > \over \sim \;$}
\def\simgt{\lower.5ex\hbox{\gtsima}} 

\def\deg{\hbox{$^\circ$}}

\def\phflux{ph\,cm$^{-2}$\,s$^{-1}$}
\def\stphflux{$\times\,10^{-6}$\,ph\,cm$^{-2}$\,s$^{-1}$}
\def\fgamma{$F_{\gamma}$}

\def\gray{$\gamma$-ray}

\def\cm3{\ifmmode{~{\rm cm^{-3}}}\else{~cm$^{-3}$}\fi} 
\def\kms{\ifmmode{~{\rm km\,s^{-1}}}\else{~km\,s$^{-1}$}\fi}

\def\r95{$r_{\rm 95}$}
\def\t0{$t_{\rm 0}$}

\def\Ms{\ifmmode{~{\rm M_\odot}}\else{M$_\odot$}\fi} 
\def\p0{$\pi^{\rm 0}$}

\def\Fermi{\textit{Fermi}}
\def\Swift{\textit{Swift}}

\def\atel{ATel} 
\def\natas{NatAs}
\def\sci{Sci} 

\usepackage{color}

\usepackage[switch]{lineno}

\usepackage{hyperref}

\accepted{July 3, 2022, ApJ}

\shorttitle{Fermi-LAT Detection of Nova RS Oph 2021}
\shortauthors{Cheung et al.}

\begin{document}

\title{\Fermi-LAT Gamma-ray Detection of the Recurrent Nova RS Ophiuchi during its 2021 Outburst}

\correspondingauthor{C.~C.~Cheung, T.~J.~Johnson, P.~Jean}
\email{Teddy.Cheung@nrl.navy.mil, Tyrel.Johnson.ctr@nrl.navy.mil, Pierre.Jean@irap.omp.eu}

\author[0000-0002-4377-0174]{C.~C.~Cheung}
\affiliation{Space Science Division, Naval Research Laboratory, Washington, DC 20375, USA}

\author[0000-0002-2771-472X]{T.~J.~Johnson}
\affiliation{College of Science, George Mason University, Fairfax, VA 22030, resident at Naval Research Laboratory, Washington, DC 20375, USA}

\author[0000-0002-1757-9560]{P.~Jean}
\affiliation{CNRS, IRAP, F-31028 Toulouse cedex 4, France} 
\affiliation{GAHEC, Universit\'e de Toulouse, UPS-OMP, IRAP, Toulouse, France}

\author[0000-0002-0893-4073]{M.~Kerr}
\affiliation{Space Science Division, Naval Research Laboratory, Washington, DC 20375, USA}

\author[0000-0001-5624-2613]{K.~L.~Page}
\affiliation{School of Physics and Astronomy, University of Leicester, Leicester LE1 7RH, UK}

\author[0000-0002-1041-7542]{J.~P.~Osborne}
\affiliation{School of Physics and Astronomy, University of Leicester, Leicester LE1 7RH, UK}

\author[0000-0001-5186-5950]{A.~P.~Beardmore}
\affiliation{School of Physics and Astronomy, University of Leicester, Leicester LE1 7RH, UK}

\author[0000-0001-5991-6863]{K.~V.~Sokolovsky}
\affiliation{Center for Data Intensive and Time Domain Astronomy, Department of Physics and Astronomy, Michigan State University, 567 Wilson Rd, East Lansing, MI 48824, USA}
\affiliation{Sternberg Astronomical Institute, Moscow State University, Universitetskii pr.~13, 119992 Moscow, Russia}

\author[0000-0002-9107-9791]{F.~Teyssier}
\affiliation{Astronomical Ring for Amateur Spectroscopy Group, 76100 Rouen, France}

\author[0000-0002-0712-2479]{S.~Ciprini}
\affiliation{Istituto Nazionale di Fisica Nucleare, Sezione di Roma ``Tor Vergata", I-00133 Roma, Italy}
\affiliation{Space Science Data Center - Agenzia Spaziale Italiana, Via del Politecnico, snc, I-00133, Roma, Italy}

\author[0000-0003-0766-6473]{G.~Mart\'i-Devesa}
\affiliation{Institut f\"ur Astro- und Teilchenphysik, Leopold-Franzens-Universit\"at Innsbruck, A-6020 Innsbruck, Austria}

\author[0000-0003-0219-4534]{I.~Mereu}
\affiliation{Dipartimento di Fisica, Universit\`a degli Studi di Perugia, I-06123 Perugia, Italy}
\affiliation{Istituto Nazionale di Fisica Nucleare, Sezione di Perugia, I-06123 Perugia, Italy}

\author[0000-0002-0130-2460]{S.~Razzaque}
\affiliation{Centre for Astro-Particle Physics (CAPP) and Department of Physics, University of Johannesburg, PO Box 524, Auckland Park 2006, South Africa}

\author[0000-0002-7376-3151]{K.~S.~Wood}
\affiliation{Praxis Inc., Alexandria, VA 22303, resident at Naval Research Laboratory, Washington, DC 20375, USA}

\author[0000-0003-1677-8004]{S.~N.~Shore}
\affiliation{Dipartimento di Fisica ``Enrico Fermi'', Universit\`a di Pisa, and INFN -- Sezione di Pisa, largo B.~Pontecorvo 3, 56127 Pisa, Italy}

\author{S.~Korotkiy}
\affiliation{Astroverty, Ka-Dar astronomy foundation, Kuzminki, P.O. Box 82, 142717 Moscow region, Russia}

\author{A.~Levina}
\affiliation{Israeli Astronomical Association, Tel Aviv, Israel}

\author{A.~Blumenzweig}
\affiliation{Israeli Astronomical Association, Tel Aviv, Israel}

\begin{abstract}

We report the \Fermi-LAT \gray\ detection of the 2021 outburst of the symbiotic recurrent nova RS~Ophiuchi. 
In this system, unlike classical novae from cataclysmic binaries, the ejecta from the white dwarf form shocks when interacting with the dense circumstellar wind environment of the red giant companion. 
We find the LAT spectra from 50\,MeV~to~$\sim$20--23\,GeV, the highest-energy photons detected in some sub-intervals, are consistent with \p0-decay emission from shocks in the ejecta as proposed by \citet{tat07} for its previous 2006 outburst.
The LAT light-curve displayed a fast rise to its peak $>$0.1\,GeV flux of $\simeq$\,6\,\stphflux\ beginning on day~0.745 after its optically-constrained eruption epoch of 2021~August~8.50. 
The peak lasted for $\sim$1\,day, and exhibited a power-law decline up to the final LAT detection on day~45.
We analyze the data on shorter timescales at early times and found evidence of an approximate doubling of emission over $\sim$200\,minutes at day~2.2, possibly indicating a localized shock-acceleration event.
Comparing the data collected by the AAVSO, we measured a constant ratio of $\sim$$2.8\,\times\,10^{-3}$ between the \gray\ and optical luminosities except for a $\sim$5$\times$ smaller ratio within the first day of the eruption likely indicating attenuation of $\gamma$~rays by ejecta material and lower high-energy proton fluxes at the earliest stages of the shock development.
The hard X-ray emission due to bremsstrahlung from shock-heated gas traced by the \Swift-XRT 2--10\,keV light-curve peaked at day~$\sim$6, later than at GeV and optical energies.
Using X-ray derived temperatures to constrain the velocity profile, we find the hadronic model reproduces the observed $>$0.1\,GeV light-curve.

\end{abstract}


\section{Introduction} 
\label{sec:intro}

RS Ophiuchi (RS Oph) is one of the best-studied recurrent novae because of its numerous outbursts since the first detection in 1898 \citep{pic05}.
It has recurred at irregular intervals \citep[with inferred outbursts missed due to Solar conjunction;][]{opp93,sch04} of 9 to 27 years up to its previous outburst on 2006 February 12 \citep{nar06,eva08,sch10}.
RS Oph is a symbiotic binary system with a 453.6\,$\pm$\,0.4 day orbital period consisting of a massive white dwarf (1.2--1.4\,\Ms) and a red giant (RG) commonly identified as type M0~III \citep{dob94,bar08,bra09}.
Its widely adopted distance of 1.6\,$\pm$\,0.3\,kpc \citep{hje86,bod87} is assumed here to facilitate direct comparison to previous work; this distance is consistent with the value of $1.4^{+0.6}_{-0.2}$\,kpc obtained considering various methods \citep{bar08} -- but see  
arguments for greater distances presented by \citet{sch09,sch18}, \citet{mon22}, and \citet{mag22}.

Its 2006 outburst was well studied \citep{eva08} with observations of hard X-ray emission from 2--25\,keV \citep{sok06} and 14--50\,keV \citep{bod06} that indicated shocked emission in the nova ejecta, and high-resolution radio imaging resolved the shocked regions \citep{obr06,rup08,sok08}. 
In $\gamma$ rays, \citet{tat07} predicted high-energy particle acceleration in the nova ejecta from interactions with the dense RG wind that could have been observed in 2006 at GeV energies\footnote{See the talk given by \citet{tat08} at \url{https://www.astro.keele.ac.uk/rsoph/pdfs/tatischeff.pdf}}, but that explosion preceded the launch of \Fermi\ in 2008.
Instead, the first $>$0.1\,GeV detection of a nova by the \Fermi\ Large Area Telescope \citep[LAT;][]{atw09} in 2010 was of a less-known symbiotic binary V407 Cyg \citep{abd10,che10}.
This discovery demonstrated the viability of the nova explosion in RG wind model and helped to solidify the predictions of GeV emission in RS Oph specifically \citep{her12}. 

Thus the next outburst of RS Oph was highly anticipated by multi-wavelength observers, particularly its $>$0.1\,GeV observation by the \Fermi-LAT.
Indeed, a new optical outburst from RS Oph was discovered in 2021, independently by A. Amorim (August 8.913), E. Muyllaert (August 8.920), and K. Geary (August 8.931); see \citet{bec21}.
The time since its 2006 detection is 15.5 years, close to its average recurrence interval of 14.7 years \citep{sch10}.
Following the \citet{gea21} announcement, we reported the independent \Fermi-LAT detection of the nova while performing its normal all-sky monitoring during the last 6-hr interval of 2021 August 8, overlapping with the optical discovery epoch \citep{che21a}\footnote{The LAT automated science processing \citep[ASP;][see Sec.~2.6.3 therein]{atw09} pipeline for these data completed on 2021 August 9, 02:02 UT and was publicly reported on August 9, 05:05 UT.}.

Here, we present the details of the \Fermi-LAT observations of RS Oph 2021.
In the following, Section~\ref{sec:lat} describes the LAT observations and analysis.
Section~\ref{sec:results} describes the LAT results, and comparison to optical and X-ray data.
Section~\ref{sec:piondecay} presents the results of the LAT spectral variability analysis for four defined emission phases and examines the originally-proposed \p0-decay model \citep{tat07,her12} to reproduce the \gray\ spectra and light-curves.
The results are summarized in Section~\ref{sec:summary}.
All times are UTC, while days relative to the optical eruption epoch, \t0\,=\,2021 August 8.5 \citep[][and \S~\ref{sec:optical}]{mun21} are used to describe the RS Oph outburst.

\begin{figure}[t]
\begin{center}
\includegraphics[scale=0.5,angle=0]{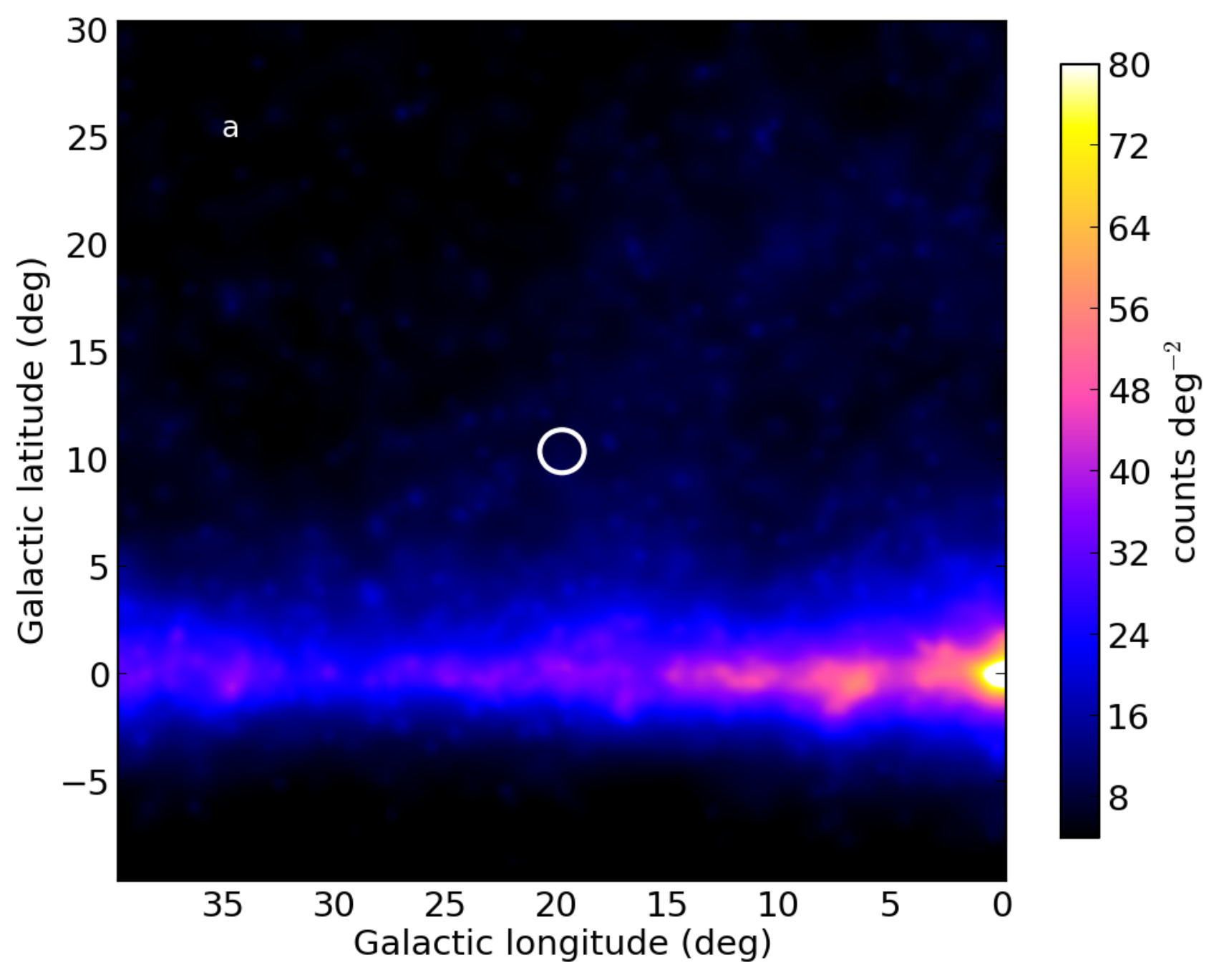}\includegraphics[scale=0.5,angle=0]{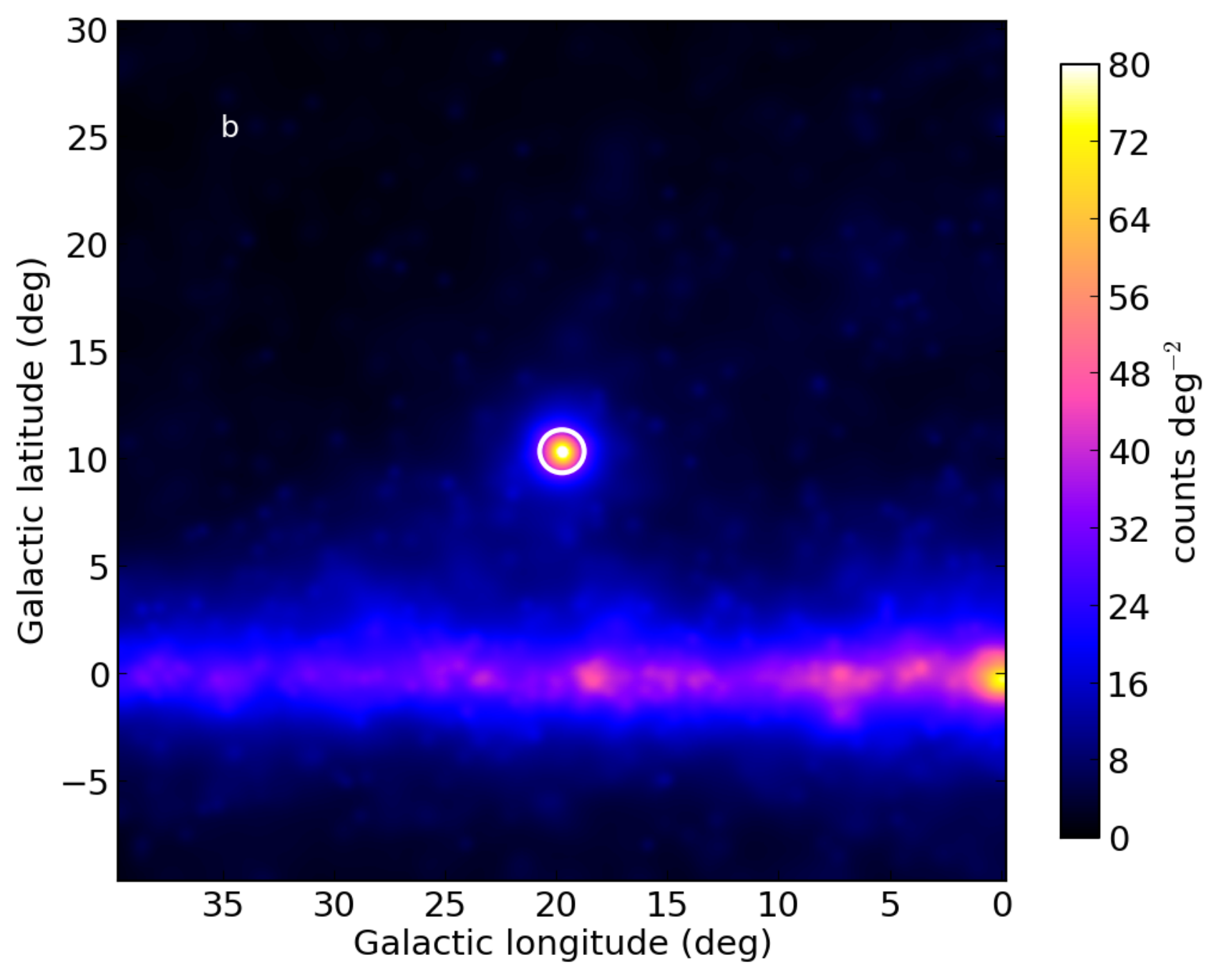}
\caption{Adaptively smoothed LAT count maps of 0.2--5\,GeV energy photons centered on the optical position of RS Oph (marked with a 1$\deg$ radius circle).
For visualization purposes, the time intervals in the panels were selected to have similar exposure before (a: July 29.0 to August 6.5) and after (b: August 9.0 to 11.8) its 2021 outburst. 
Note the structured Galactic diffuse emission in the bottom portion of the images.}
\label{fig:countsmap}
\end{center}
\end{figure}

\section{\Fermi-LAT Observations and Data Analysis}
\label{sec:lat}

For the LAT analysis, we used Pass 8 (P8R3) \texttt{SOURCE} class data\footnote{See \url{https://fermi.gsfc.nasa.gov/ssc/data/analysis/documentation/Pass8_usage.html}} \citep{atw13,bru18}, as defined under the \texttt{P8R3\_SOURCE\_V3} instrument response functions. 
Photons with 0.05--300\,GeV energies, within 25\deg\ of  R.A., Decl.~(J2000) = $267\fdg7$, $-6\fdg7$, and with maximum zenith angle of 90\deg\ were selected.
The center of the region of interest (ROI) was chosen to put RS Oph near the center but not on a pixel edge in the binned analysis described below.
We filtered the events to include only good time intervals when the LAT data were flagged as good and the instrument was in nominal science observations mode.  
All data processing and analysis was done using version 2.0.8 of \texttt{fermitools} \citep{fer19}.

We constructed a spatial and spectral model of our ROI starting from the 4FGL catalog incremental data release 3 \citep[DR3;][]{abd20,abd22} based on twelve years of LAT data. 
We included all DR3 sources within 35\deg\ of the ROI center as well as a model for the Galactic diffuse emission (\texttt{gll\_iem\_v07.fits}) and a diffuse isotropic emission component (\texttt{iso\_P8R3\_SOURCE\_V3\_v1.txt})\footnote{Both files are available for download, \url{https://fermi.gsfc.nasa.gov/ssc/data/access/lat/BackgroundModels.html}}.
Note the source of emission from RS Oph as seen in the counts map (Figure~\ref{fig:countsmap}) is relatively isolated from other point sources (the closest 4FGL-DR3 source is 4FGL~J1752.4--0758, offset by 1\fdg38) and the Galactic diffuse emission, but is close enough to the latter to potentially cause low-level contamination at the nova position, especially at the lowest energies.

We added a point-source to our model at the optical position of the nova (R.A., Decl.~= $267\fdg5550$, $-6\fdg7079$), initially assuming a single power-law (PL) spectral shape, $(dN/dE)\,=\,N_{0}\,(E/E_{0})^{-\Gamma}$, and fitted the normalization ($N_{0}$) and photon index ($\Gamma$), while keeping the scale energy fixed to $E_{\rm 0}$\,=\,1\,GeV.
We performed a binned maximum likelihood analysis on a $35\fdg3\times35\fdg3$ square region, binning data into $0\fdg1\times0\fdg1$ pixels.
For our starting model, the spectral parameters of sources within 15\deg\ of the ROI center were allowed to vary if they were found in the 4FGL-DR3 analysis with test statistic \citep{mat96}, TS\,$\geq$\,100, while the parameters of all other sources were held fixed.  
We left the normalization and index of the Galactic diffuse spectrum free in the fit, and the isotropic component also had a free normalization.  
To refine our starting model for the ROI before analyzing the outburst, we fit a one-year dataset spanning 2020 July 19 to 2021 July 19, which ends 20 days before \t0.
The fit was done over the energy range 0.05--300\,GeV, including the effect of energy dispersion in our analysis (except for the isotropic component), and resulted in no significant detection at the position of the nova with a 95\% confidence flux upper limit, $<$$5.3\times10^{-8}$\,\phflux\ ($<$$1.5\times10^{-8}$\,\phflux\ from 0.1--300\,GeV).
Using the results of this fit, we created a new model of the region for the following 10-day spectral analyses of the nova outburst.
In this updated model, we fixed the spectral parameters of all the point sources in the ROI that had been free but had TS\,$<$\,100 in the one-year analysis.  
For point sources with free parameters found to have TS\,$\geq$\,100, we left the normalization parameters free but fixed all other spectral parameters.  
We also fixed the spectral parameters of all extended sources and the Galactic diffuse emission while leaving the normalization of the isotropic component free to vary.

Analyses of previous \gray\ novae found significant curvature in the spectrum \citep[e.g.,][]{ack14} with the best-fit spectra using an exponentially cutoff power-law (ECPL) shape, $(dN/dE)\,=\,N_{0}\,(E/E_{0})^{-\Gamma}\,\exp(-aE)$, where $a$ is the fitted exponential factor\footnote{This functional form of an ECPL is more stable than other available models in \url{https://fermi.gsfc.nasa.gov/ssc/data/analysis/scitools/source_models.html}, and a cutoff energy can be derived as $E_{\rm cut}\,=\,a^{-1}$.}.
To test for curvature in the \gray\ spectrum of RS Oph, we analyzed the time period from 2021 August 8.5 to 18.5, corresponding to the main activity when there were consecutive power-law fluxes, $F_{\gamma}$\,($>$0.1\,GeV)\,$\geq$\,1\,\stphflux\ based on preliminary results \citep{che21b}\footnote{Following the initial LAT detection of RS Oph on August 8.75–9.00 (\S~1), we continued to monitor it through an automated daily analysis pipeline started by the LAT team in March 2021 aimed at detecting the anticipated nova outburst independent of other instruments. The main differences between this preliminary analysis and the final results presented are the background source model (4FGL), the time range of the dataset used to fit the background, and a single PL spectral model was adopted for RS Oph throughout.}.
We performed a binned maximum likelihood analysis on this 10-day period over the 0.05--300\,GeV energy range, similar to what was done for the 1-year period prior to the outburst and starting from the updated model resulting from that fit.
The nova was modeled using both an ECPL and a single PL, deriving TS$_{\rm curve}\,=\,-2(\mathcal{L}_{\rm ECPL}\,-\,\mathcal{L}_{\rm PL})$ as a measure of significance of curvature in the spectrum, where $\mathcal{L}$ are the respective best-fit $-$log likelihood values of the two fits.  
The results show $\sim$8$\sigma$ (TS$_{\rm curve}$\,=\,66) evidence for curvature in the spectrum with best-fit ECPL parameters (TS\,=\,2856), $\Gamma\,=1.66\,\pm\,0.05$ and $E_{\rm cut}\,=\,6.0\,\pm\,1.2$\,GeV, with 0.05--300\,GeV and 0.1--300\,GeV fluxes = (3.60\,$\pm$\,0.20)\,\stphflux\ and (2.16\,$\pm$\,0.09)\,\stphflux, respectively.
A significant detection in the lowest-energy bin from 50--100\,MeV (TS\,=22, or 4.7$\sigma$) helped to constrain the curvature.
The spectral curvature is likely due to the hadronic origin of the emission (see \S~\ref{sec:piondecay}).

Though the correlated variability provides a firm identification of the LAT source with the nova, we used the \texttt{fermitools} \texttt{gtfindsrc} to localize the \gray\ emission during the first 10-day main activity interval. 
We selected 1--300\,GeV photons because these events have the best per-photon resolution \citep{abd20} while also excluding the energy range where the Galactic diffuse emission is dominant.
The resulting R.A., Decl. = ($267\fdg558,-6\fdg736$) is offset by 0\fdg028 from the optical position, and within the 95\% confidence-level containment radius of $r_{\rm 95}\,=\,0\fdg029$ (statistical only).

\subsection{LAT light-curves}
\label{sec:lightcurves}

We generated \gray\ flux light-curves starting 20 days before the optical eruption epoch, \t0\,=\,2021 August 8.5 \citep[][see \S~\ref{sec:optical}]{mun21} and ending at \t0+92 days in 6-hr, 1-day, and 4-day bins.
The end date of day 92 (2021 November 8.5) probes $\sim$$2\times$ further than the last significant ($\geq$3$\sigma$) LAT detection at day $\sim$45 (see below) and coincides with the end of the \Swift\ observing season (\S~\ref{sec:swift}).  
We produced light-curves at $>$0.1\,GeV energies to facilitate comparison to previous LAT-detected novae \citep[e.g.,][]{ack14,che16}, thus we restricted the events to 0.1 to 300\,GeV photons within 15\deg\ of the ROI center for these analyses.
In each time-bin, we performed a binned maximum likelihood analysis on a $21\fdg2\times21\fdg2$ region, binned into pixels $0\fdg1$ on a side.
We adopted the best-fit background model from the analysis of the 10-day dataset described in \S~\ref{sec:lat}, while freezing the spectral normalizations of the field point sources with TS\,$<$\,80 in that analysis, and removed sources more than 25\deg\ from the ROI center.  
For the nova, we used an ECPL model with a fixed $a$ parameter (corresponding to the best-fit $E_{\rm cut}$\,=\,6\,GeV from the 10-day interval) while fitting the normalization and the photon index.
The resultant LAT 6-hr, 1-day, and 4-day bin $>$0.1\,GeV flux light-curves and fitted photon indices are presented in the Appendix (\S~\ref{sec:appendixhe}).

For the first three days of the outburst, we performed further analyses on individual \Fermi\ spacecraft orbits ($\sim$1.6\,hrs).
The orbit-binned analysis more accurately reflects the details of the LAT exposure profile at the RS Oph position at early times when the source was brightest (see Figure~\ref{fig:gammaopticalxray}, top panel).
Specifically, each of the analyzed orbital intervals consist of $\sim$1 hr ($\sim$0.04 days) of exposure every $\sim$3.2 hrs (see \S~\ref{sec:appendixorbit} for details).

Composite results from the different time-binned results are shown Figure~\ref{fig:gammaopticalxray} and Figure~\ref{fig:gammaplus}.
Throughout, we consider significant detections at TS\,$\geq$\,12 ($\geq$3$\sigma$ significance for two degrees of freedom, d.o.f.), as indicated with black points in each LAT data panel.
To help visualize fluxes at lower-significance, particularly in the 4-day analysis, we use gray points with error bars to indicate intervals with TS\,=\,6--12 (2--3$\sigma$). 
We report 95\% confidence-level flux upper limits for time-bins in which the nova was found with TS\,$<$\,6 ($<$2$\sigma$), less than 4 predicted counts ($N_{\rm pred}$), or had determined uncertainties greater than the fitted fluxes.
The results derived from the light-curves and spectral analysis of four defined \gray\ emission phases are given in \S~\ref{sec:results} and \S~\ref{sec:piondecay}, respectively.

\subsection{Shortest-timescale LAT analysis}
\label{sec:shortest}

We searched for variability on the shortest timescales in the first 10 days of LAT data by applying the methods of \citet{ker19} to estimate the likelihood as a function of only the nova flux in 10-minute time-bins (Figure~\ref{fig:gammaplus}, top panel).
In this analysis, relative fluxes, $F_{\gamma, \rm rel}$ (relative to the mean flux over the first 10-days set at 1.0) with error bars are shown for points with TS\,$\geq$\,4 corresponding to $\geq$\,2$\sigma$ significances (significance $\sim\sqrt{{\rm TS}}$ for 1 d.o.f.).
We subsequently grouped these intervals into longer partitions with the Bayesian Blocks (BB) algorithms \citep{sca13}.  
Additionally, we estimated the likelihood for each spacecraft orbit-bin (not shown) and confirmed the results of the \texttt{gtlike} analysis reported above.

The 10-minute analysis indicates a potential additional $\sim$20-minute duration feature at day 2.21 with $F_{\gamma, \rm rel}$\,=\,3.45\,$\pm$\,0.54.
This is $>$2$\times$ brighter than the measurements on the preceding day 1.94--2.20 and subsequent day 2.22--3.80 bins with respective $F_{\gamma, \rm rel}$\,=\,1.35$\,\pm$\,0.21 and 1.50\,$\pm$\,0.10.
This feature also appears in the orbit-binned analysis at day 2.207 (60-minute bin), albeit at a slightly less pronounced level, with a flux, $F_{\gamma}$($>$0.1\,GeV)\,=\,(5.8\,$\pm$\,1.2)\,\stphflux, while the adjacent orbit-bin fluxes are $\simeq$\,(3.4\,$\pm$\,1.0)\,\stphflux, amounting to $\sim$1.5$\sigma$ differences.
Assessing the significance of features obtained with the BB algorithm is challenging.  
There are 431 individual 10-minute intervals in the analyzed range, and we adopted an exponential prior on the number of change points $\pi(N_{p})^{-\gamma}$ with $\gamma=7$, yielding a ``false positive rate'' for a spurious change point of 0.4.  
However, if we restrict attention to the first three days when variations can be more easily detected when the source was brightest, the false positive rate drops to $\sim$0.1.  
Thus we estimate the total significance of this $\sim$20-minute duration feature at day 2.21 (involving two additional change points) to be about 2$\sigma$.

We additionally split select orbits into two bins and performed the \texttt{gtlike} analysis (see \S~\ref{sec:appendixorbit}).
In only one orbit bin was there observed flux variability between the two 30-minute intervals, indicating an observed peak flux, $F_{\gamma}$($>$0.1\,GeV)\,=\,(8.1\,$\pm$\,1.8)\,\stphflux\ at day 1.40, halving to (4.0\,$\pm$\,1.2)\,\stphflux\ ($\sim$2$\sigma$ difference).
This feature was not observed in the BB analysis above.

\subsection{Highest-energy LAT photons}
\label{sec:he}

We searched for the highest-energy LAT events from \t0--20 to +92 days by selecting $>$5\,GeV photons within 0\fdg5 from the optical position of RS Oph.
The resultant list of photons is given in the Supplement (\S~\ref{sec:appendixhe}).
The times and energies of the photons detected during the first 10-day main activity interval are shown in Figure~\ref{fig:gammaplus}.

There are eight $>$10\,GeV LAT photons detected with \texttt{gtsrcprob} weight value (probability of association calculated using the best-fit ECPL model for the 10-day dataset described in \S~\ref{sec:lat}) of $>$\,0.90 observed within 10 days of \t0.
Curiously, the first of these events (10.5\,GeV) at day 0.222 is in the orbit-bin prior to the first orbit detection on day 0.352 ($\pm$0.023). 
The highest energy photon is 23\,GeV on day 2.861, just after the flux peak.

At later times, after the main 10-day activity interval (see the figure in \S~\ref{sec:appendixhe}), the highest-energy photons detected with \texttt{gtsrcprob} weight values $>$\,0.90 were on days 13.8 and 19.2 ($E$ = 11 and 19 GeV, respectively).
Outside the 45-day LAT emission duration (\S~\ref{sec:lightcurves}), a single 12.8\,GeV photon with smaller \texttt{gtsrcprob} weight value = 0.85 was found at day 55.

\begin{figure}[t]
\begin{center}
\includegraphics[scale=0.75,angle=0]{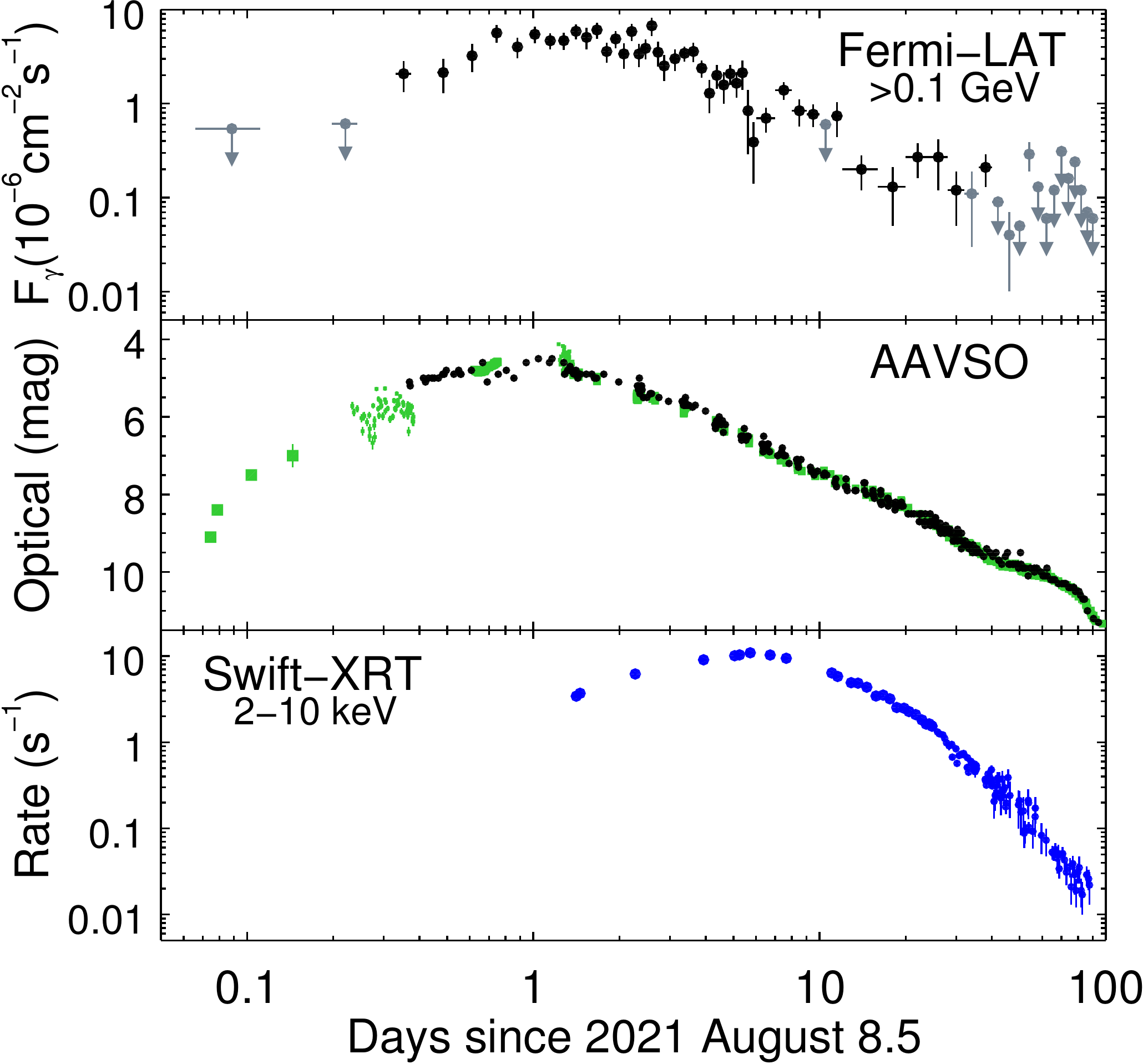}
\caption{LAT \gray\ (top), AAVSO optical (middle), and \Swift-XRT (bottom) light-curves of RS Oph 2021 relative to \t0.
The LAT light-curve shows various binning as follows: orbit-timescale with typical exposure of 60 minutes up to day 3.0, then 6-hr up to day 6.0, 1-day up to day 12.0, and 4-day thereafter; 
data with error bars (black) are $>$3$\sigma$ detections, while 2--3$\sigma$ data and upper limits are shown in gray. 
The optical data are $V$-band (green) and Visual observations (black) from the AAVSO; to reduce scatter, the Visual points from day 1.3 onward are 5-point median values.
Additional $V$-band measurements from days 0.233--0.381 and 1.227--1.375 are from observations described in \S~\ref{sec:optical}.
The \Swift-XRT data are in the 2--10\,keV band.
}
\label{fig:gammaopticalxray}
\end{center}
\end{figure}

\begin{figure}[t]
\begin{center}
\includegraphics[scale=0.7,angle=0]{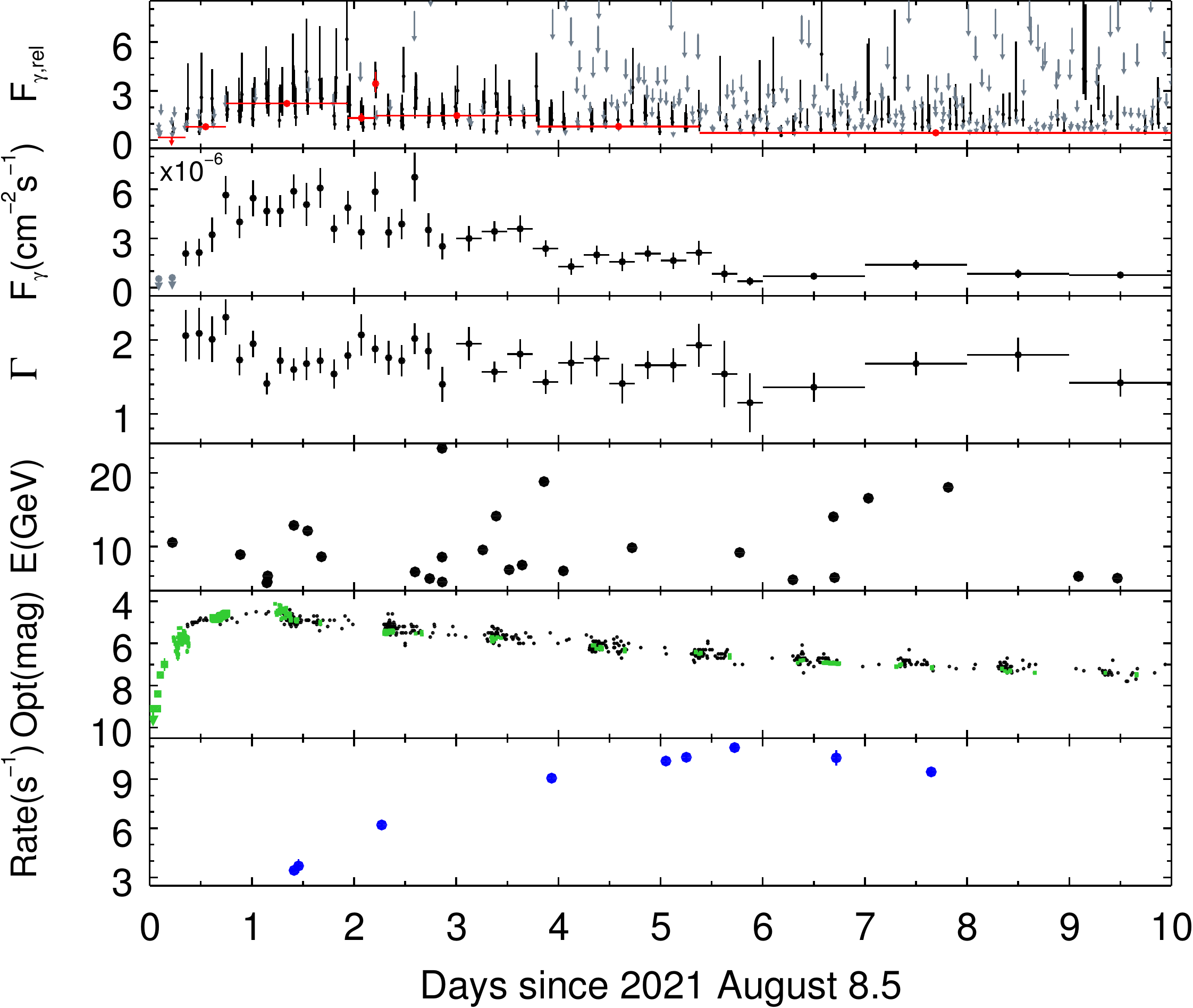}
\caption{Selected data for the first 10 days of activity for RS Oph 2021. 
From top: 
LAT $>$0.1\,GeV light-curve in 10-minute bins and units of relative flux (to the mean flux over the first 10-days, set to 1.0) with upper limits indicated with gray arrows, and Bayesian Block partitions indicated in red, 
LAT $>$0.1\,GeV composite light-curve (see Figure~\ref{fig:gammaopticalxray}, top), 
LAT photon indices, $\Gamma$,
energies of the detected $E\,>$5\,GeV photons, 
optical light-curve (see Figure~\ref{fig:gammaopticalxray}, middle; except all individual Visual observations are shown here), and
X-ray (2--10\,keV) light-curve (see Figure~\ref{fig:gammaopticalxray}, bottom).
}
\label{fig:gammaplus}
\end{center}
\end{figure}

\section{Fermi-LAT and Multiwavelength Results}
\label{sec:results}

The LAT light-curve for RS Oph 2021 shown in Figure~\ref{fig:gammaopticalxray} (top panel) is a composite of the orbit-binned, 6-hr, 1-day, and 4-day analysis relative to \t0, choosing increasing size time-bins at later times.
The LAT 6-hr and 1-day analyses show significant consecutive detections up to day 10. 
The longer, 4-day integration increased the sensitivity to lower fluxes and helped to define the light-curve at later times, when the shorter-timescale analysis resulted in more intermittent and lower-significance detections.

The main LAT-based results determined from the LAT $>$0.1\,GeV light-curve are as follows: 
(a) the observed \gray\ onset is constrained to the orbit centered on day 0.35, with a flux, \fgamma\,=\,(2.1\,$\pm$\,0.7)\,\stphflux; 
(b) the source rises to a peak \fgamma\,=\,(5.7\,$\pm$\,1.2)\,\stphflux\ at day 0.745; 
(c) the peak is flat, with an approximately constant (slope of $-0.18\,\pm\,0.17$), and average \fgamma\,$\simeq$\,5\,\stphflux\ through to day 2.5; 
(d) the flux declined by a factor of two from the day 0.745 flux value at approximately day 3;
(e) considering the last significant 6-hr detection on day 45.125 and the first orbital-bin detection at day 0.352, the total \gray\ duration was approximately 45 days; and
(f) taking the power-law slope of $-1.53\,\pm\,0.11$ fit from day 2.5 to 45, and the average peak flux from day 0.745--2.5, the source declined by a factor of 10 on day 10.

Results from additional LAT analysis performed -- the BB analysis, the spectral photon indices, and the highest-energy ($E > 5$\,GeV) photons -- are also presented in Figure~\ref{fig:gammaplus} for the main 10-day activity phase.
The above results (a) and (b) are consistent with the BB analysis with the first BB detection at day 0.350, the peak from 0.746--1.940, and the subsequent flux decline. 
The fitted LAT spectra have a wide range of spectral slopes, $\Gamma$ = 1.5 to 2.1, with typical errors of 0.2.
In the rising portion of the light-curve, the \gray\ spectral slopes were $\Gamma\,\sim$\,2.0--2.3, while the spectra hardened ($\Gamma\,\sim\,1.7$) during the peak. 
The hardest spectrum ($\Gamma\,=\,1.4\,\pm\,0.2$) was observed on day 2.87 when the highest-energy LAT photon with $E\,=$\,23.3\,GeV was detected (along with an 8.6\,GeV photon only 42.4\,s later).

To compare to the LAT data, optical and X-ray (2--10\,keV) light-curves are shown in Figure~\ref{fig:gammaopticalxray} for the entire $\sim$3-month interval studied while Figure~\ref{fig:gammaplus} details the main activity interval during the first 10 days. 
Overall, the \gray\ and optical light-curves are similar (see also Figure~\ref{fig:gammaplus}), both peaking early (around day 1; see below), while the X-rays peak later (day $\sim$6).

The optical data (\S~\ref{sec:optical}) were mainly taken from the American Association of Variable Star Observers (AAVSO) database, with additional serendipitous measurements on the first two nights (August 8, 9) obtained by a Global Meteor Network \citep{vid21} camera IL0003 (by A.L.~and A.B.) newly presented here.
During the rise in \gray\ flux, the optical increased by four magnitudes in $V$-band from observations spanning about days 0.07 to 0.80.
The power-law slope fitted to the earliest \gray\ data up to day 0.80 ($1.58\,\pm\,0.69$) is consistent with the optical $V$-band one ($1.28\,\pm\,0.01$).
The \gray\ onset observed on 2021 August 8.852 was delayed by $\sim$0.35\,day after the optical eruption (\t0), but earlier than the Visual discovery epochs by about 0.08 day.
The optically-brightest emission was observed from day 0.9--1.3 (Visual\,=\,4.5--4.6\,mag), within the wider timespan of the observed \gray\ peak fluxes from day 0.75--1.67 (see Table~\ref{table:orbit}).
Fitting a broken power-law to the Visual measurements from days 0.37--3.0, the best-fit peak was at day 1.09 ($\pm$0.12), consistent with the day 1.08 ($\pm$0.05) estimated by \citet{mun21}. 
Although the \gray\ peak fluxes appear constant over a $\sim$1-day interval, assuming a broken power-law parameterization of the composite LAT light-curve (Figure~\ref{fig:gammaopticalxray}) gives a best-fit peak at day 1.64 ($\pm$0.11) that is formally delayed with respect to the optical peak.
The slopes of the \gray\ and optical declines, $-1.35\,\pm\,0.07$ and $-1.395\,\pm\,0.006$ respectively, are consistent given the uncertainties.

\begin{figure}[t]
\begin{center}
\includegraphics[scale=0.75,angle=0]{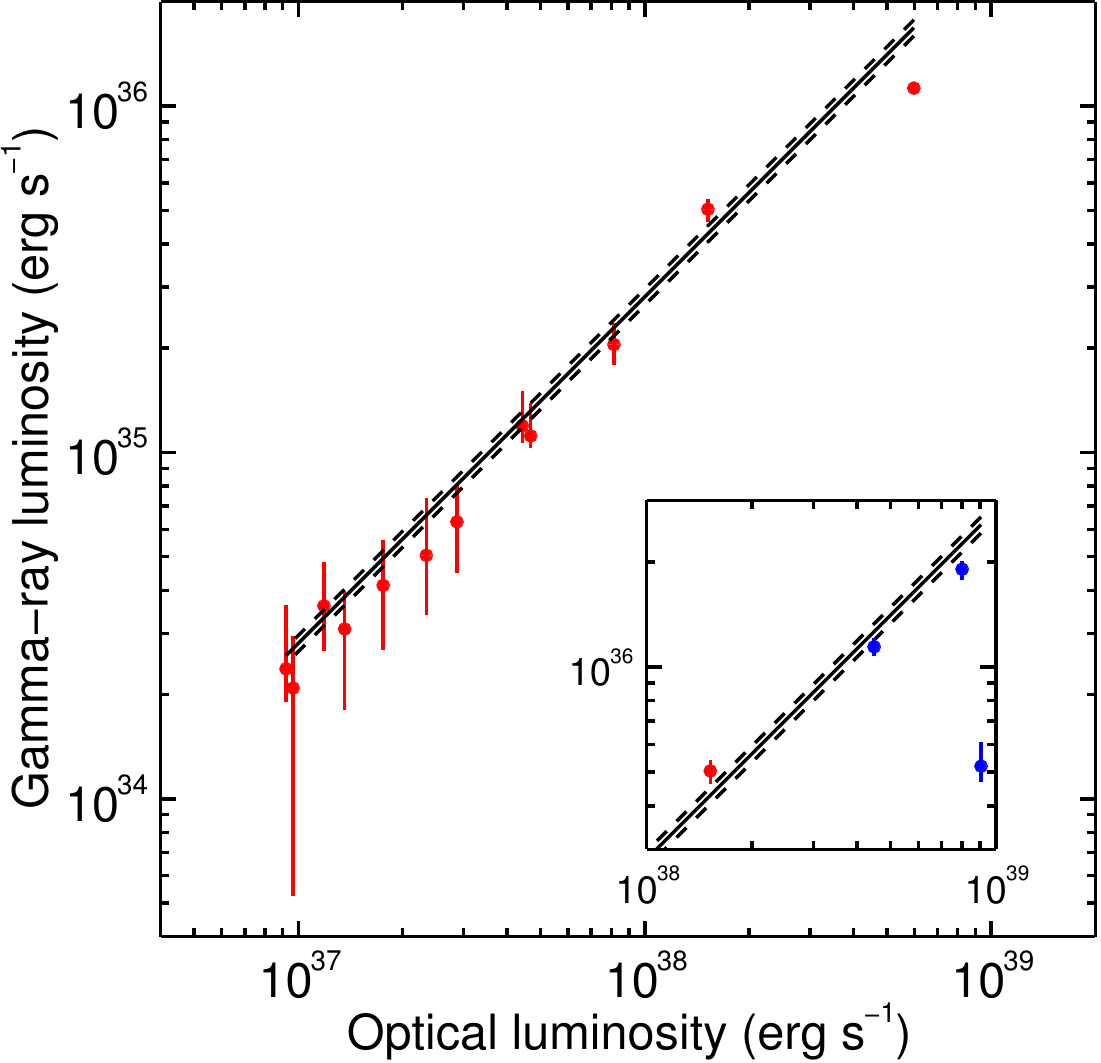}\hspace{0.05in}\includegraphics[scale=0.75,angle=0]{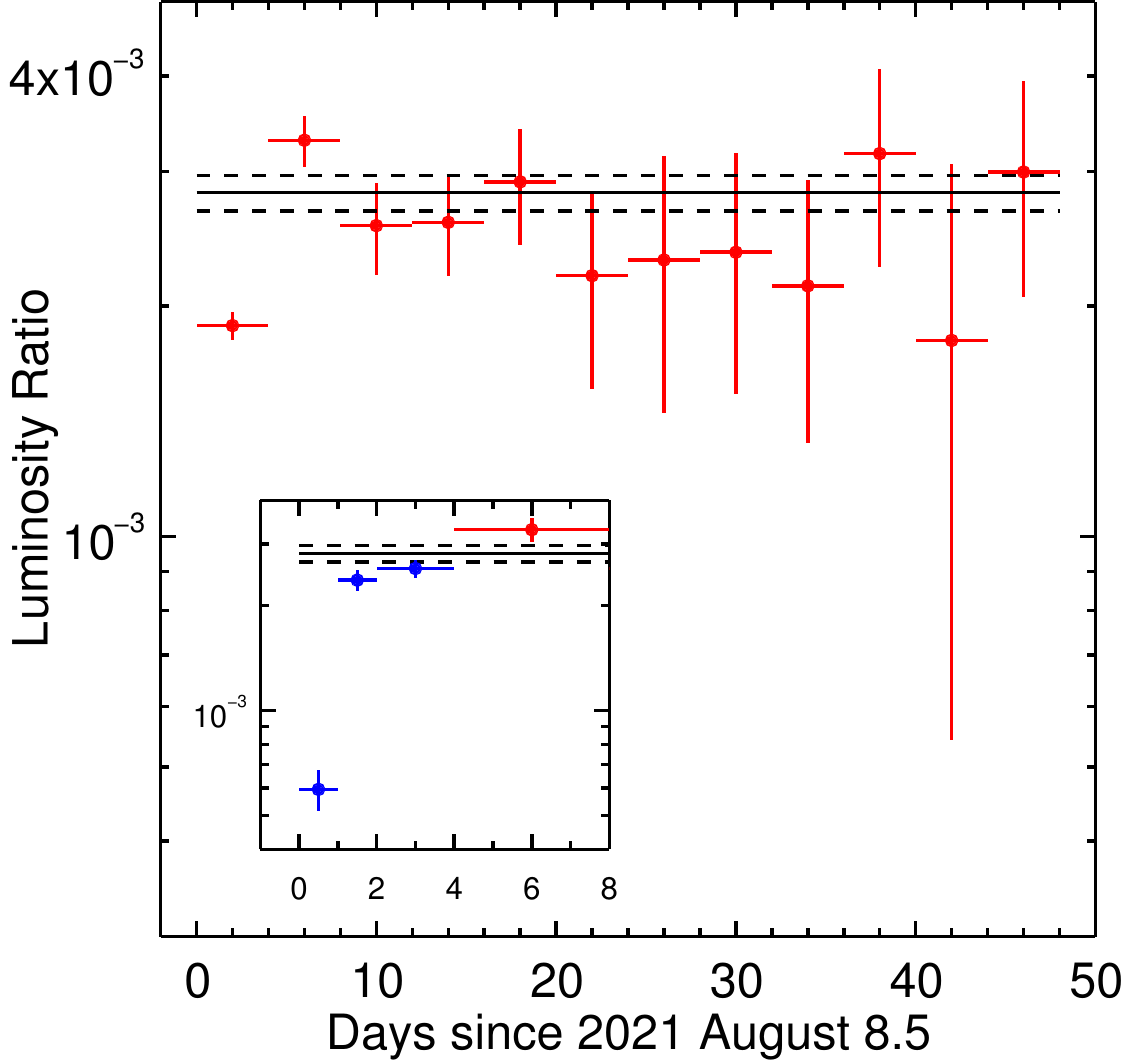}
\caption{(Left): Gamma-ray luminosity as a function of the observed optical luminosity (see \S~\ref{sec:opticallum}) for RS Oph 2021 extracted in 4-day time-bins.
The black solid line shows the best-fit ratio with its uncertainty (dashed lines) obtained by excluding the largest luminosity values (see text).
(Right): Ratio of the \gray\ luminosity to the observed optical luminosity extracted in 4-day time-bins.
The  black solid line shows the best-fit ratio, $(2.81\,\pm\,0.15)\,\times\,10^{-3}$, with its uncertainty (dashed lines),  obtained without the ratio in the bin at day 0--4.
The insets in each panel show the results from reanalysis of the highest-luminosity, earliest 4-day bin in smaller time intervals (blue data points; see text).
}
\label{fig:gammaoptical}
\end{center}
\end{figure}

We examined the best-fit \gray\ luminosity in 4-day bins obtained from the LAT data analysis\footnote{The \gray\ luminosities were calculated by fitting the fluxes and the proton spectrum slopes (without energy cutoff) in the \p0\ model, to the \Fermi-LAT data (see \S~\ref{sec:piondecay}).} as a function of the observed optical luminosity estimated in 4-day bins using the AAVSO data (see \S~\ref{sec:opticallum}).
Both the luminosities are proportional to each other, except for the largest luminosity value (Figure~\ref{fig:gammaoptical}, left).
The ratios of the \gray\ to optical luminosities in RS Oph (Figure~\ref{fig:gammaoptical}, right) are similar to those derived for classical novae \citep{met15,li17,ayd20}. 
Excluding the first 4-day bin, we calculated a luminosity ratio $L_{\gamma}(E > {\rm 50\,MeV})/L_{\rm opt}\,=\,(2.81\,\pm\,0.15)\,\times\,10^{-3}$, while the ratio in the first 4-day bin is lower by a factor of $\sim$1.5.
Reanalysis of the first 4-day bin in smaller time intervals (day 0--1, 1--2, and 2--4; blue data points in panel insets in Figure~\ref{fig:gammaoptical}) found a ratio lower by a factor of $\sim$5 in the first 1-day bin (see \S~\ref{sec:piondecay} for a discussion), while the other ratio values are compatible with the constant value.

The X-ray observations obtained with the X-ray Telescope \citep[XRT;][]{bur05} on the Neil Gehrels \Swift\ Observatory \citep{geh04} are summarized in \S~\ref{sec:swift}, and their detailed analysis is presented elsewhere \citep{pag22}.
The X-ray light-curve for the 2021 outburst is overall similar to that seen following the 2006 eruption \citep[see][]{bod06,osb11}, except that the 2006 data count rate maximum was about double that measured in 2021 \citep{pag22}. 
Here, the XRT 2--10\,keV light-curves are used to compare to the LAT \gray\ and optical light-curves.
The early X-ray emission at $>$2\,keV energies is dominated by bremsstrahlung from shocked gas in the nova ejecta \citep[e.g.,][]{sok06}, and the results of the X-ray temperature spectral fits of the 2021 XRT data presented by \citet{pag22} are used to constrain the temporal evolution of the ejecta velocity in our modeling (see \S~\ref{sec:piondecay} and Appendix~\ref{sec:swift}).
The main feature of the 2021 XRT 2--10\,keV light-curve is that the peak at day $6.4\,\pm\,0.1$ is consistent with the approximate time of the break in the X-ray derived velocity profile (\S~\ref{sec:piondecay}). 
Thereafter, the 2--10\,keV light-curve decline can be parameterized as a broken power-law with a slope = $-1.24\,\pm\,0.03$ to day 19.1 ($\pm$0.2) and a steeper decline (slope = $-3.11\,\pm\,0.03$) to the last data at day 87.6.

\section{Pion-decay Gamma-ray Emission} 
\label{sec:piondecay}

\begin{figure}[t]
\begin{center}
\includegraphics[scale=0.75,angle=0]{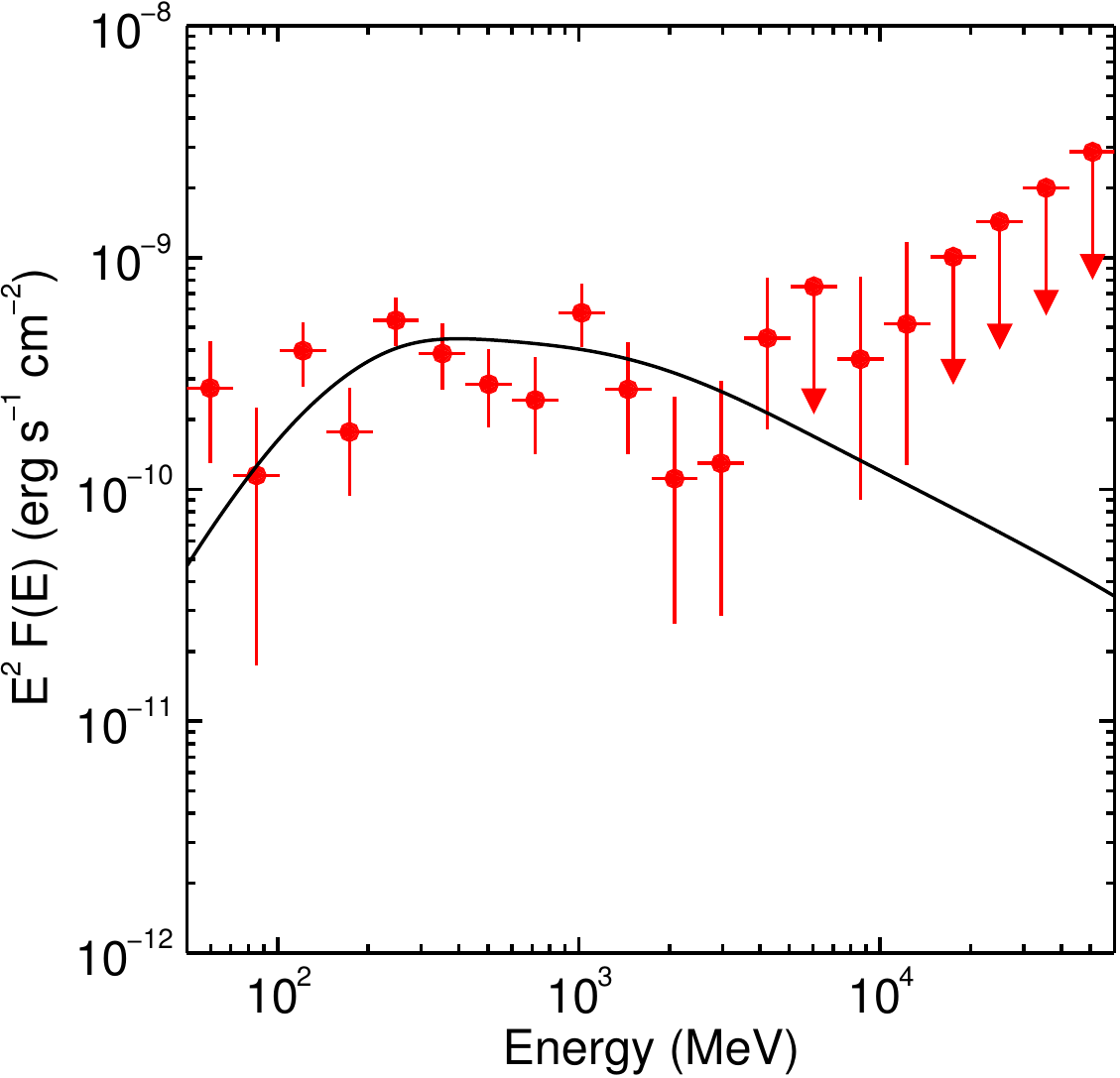}\hspace{0.1in}\includegraphics[scale=0.75,angle=0]{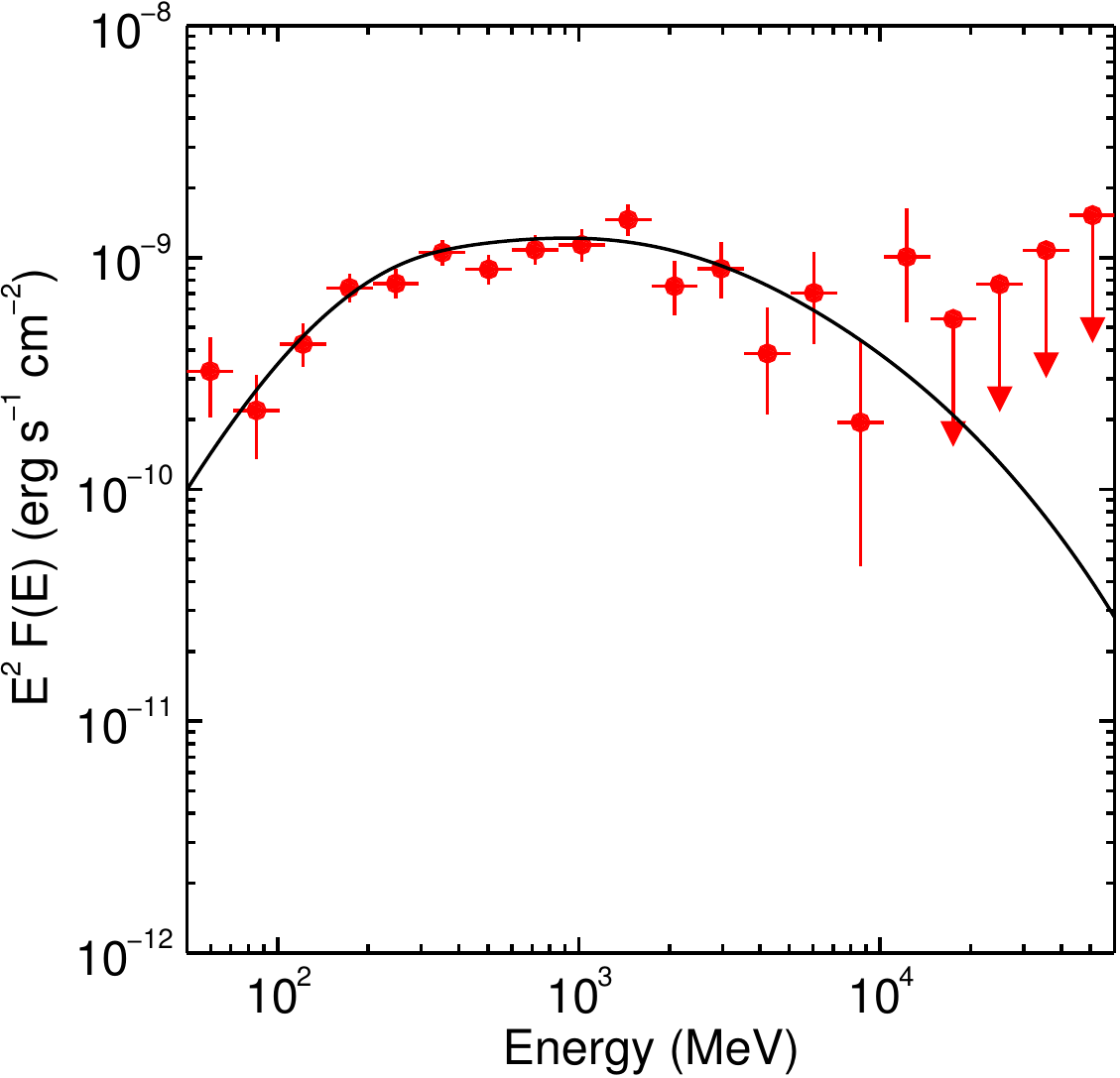} 
\includegraphics[scale=0.75,angle=0]{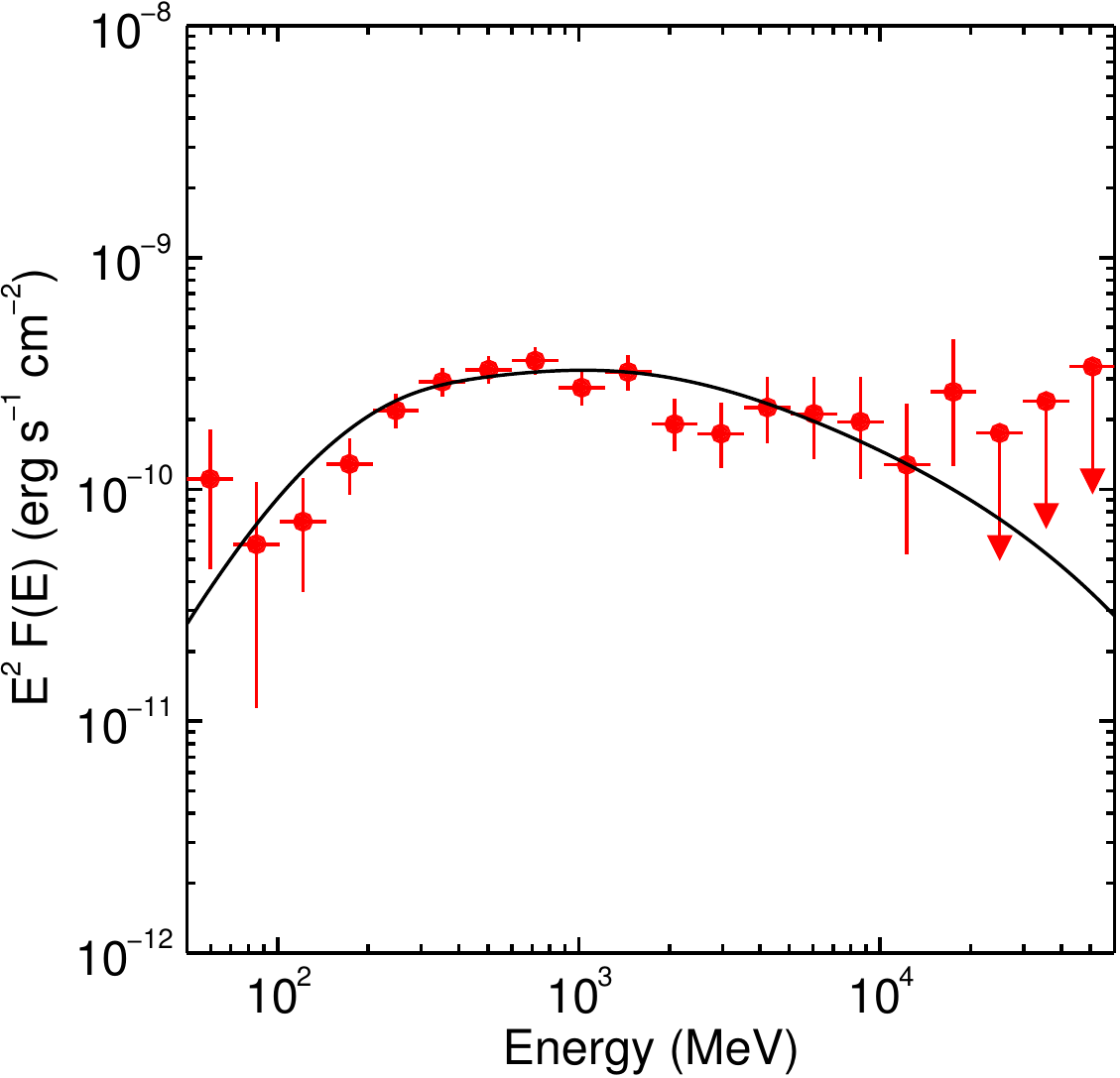}\hspace{0.1in}\includegraphics[scale=0.75,angle=0]{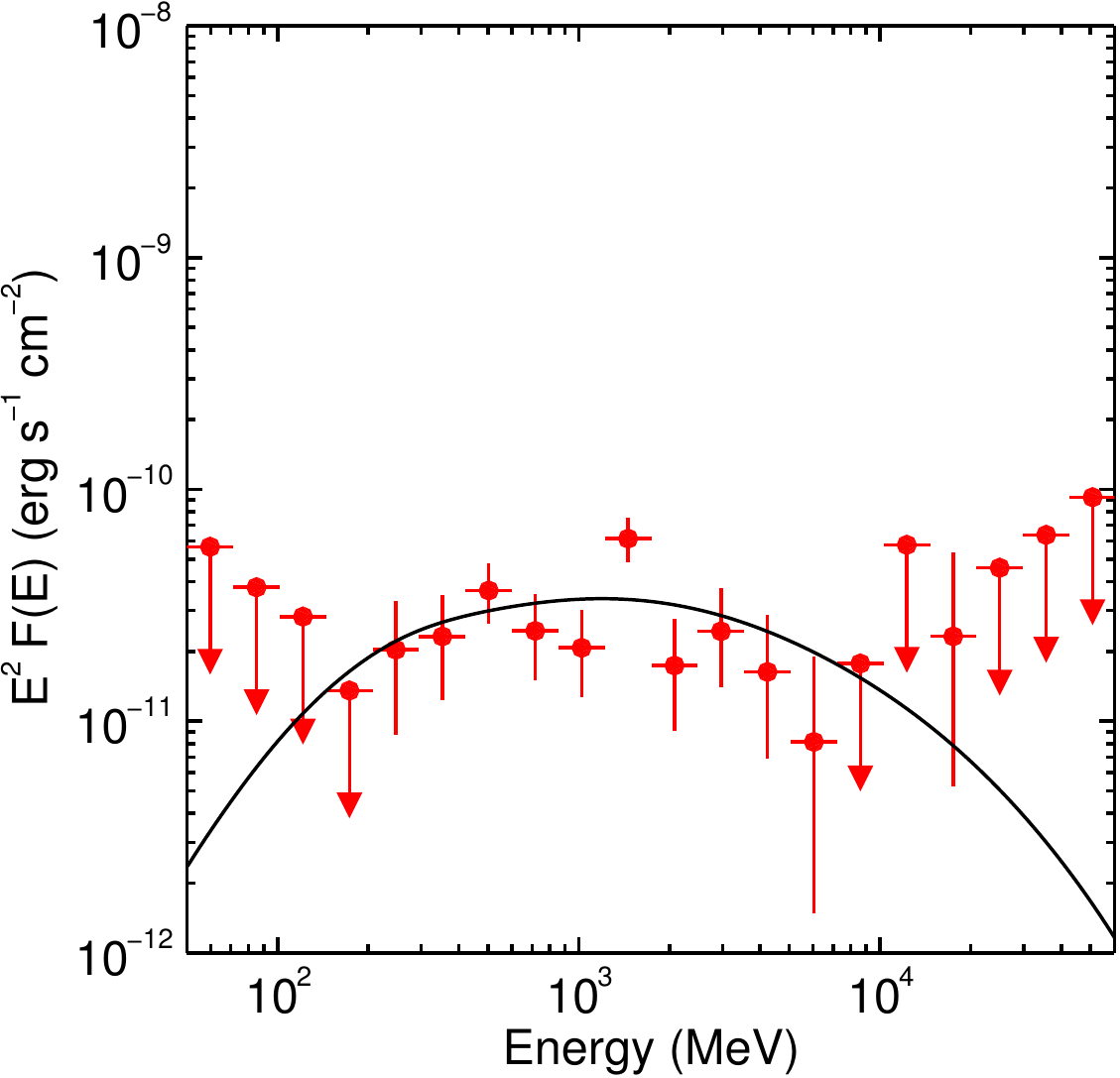} 
\caption{LAT \gray\ spectra and best-fit hadronic models for the Rise (top left), Peak (top right), Decline-a (bottom left) and Decline-b (bottom right) phases for RS Oph 2021 -- see Table~\ref{table:spectra4intervals}.
Vertical lines indicate $1\sigma$ uncertainties when TS\,$\geq$\,4.0 and arrows indicate $2\sigma$ upper limits when TS\,$<$\,4.0.} 
\label{fig:pi0spectra}
\end{center}
\end{figure}

\begin{figure}[t]
\begin{center}
\includegraphics[scale=0.79,angle=0]{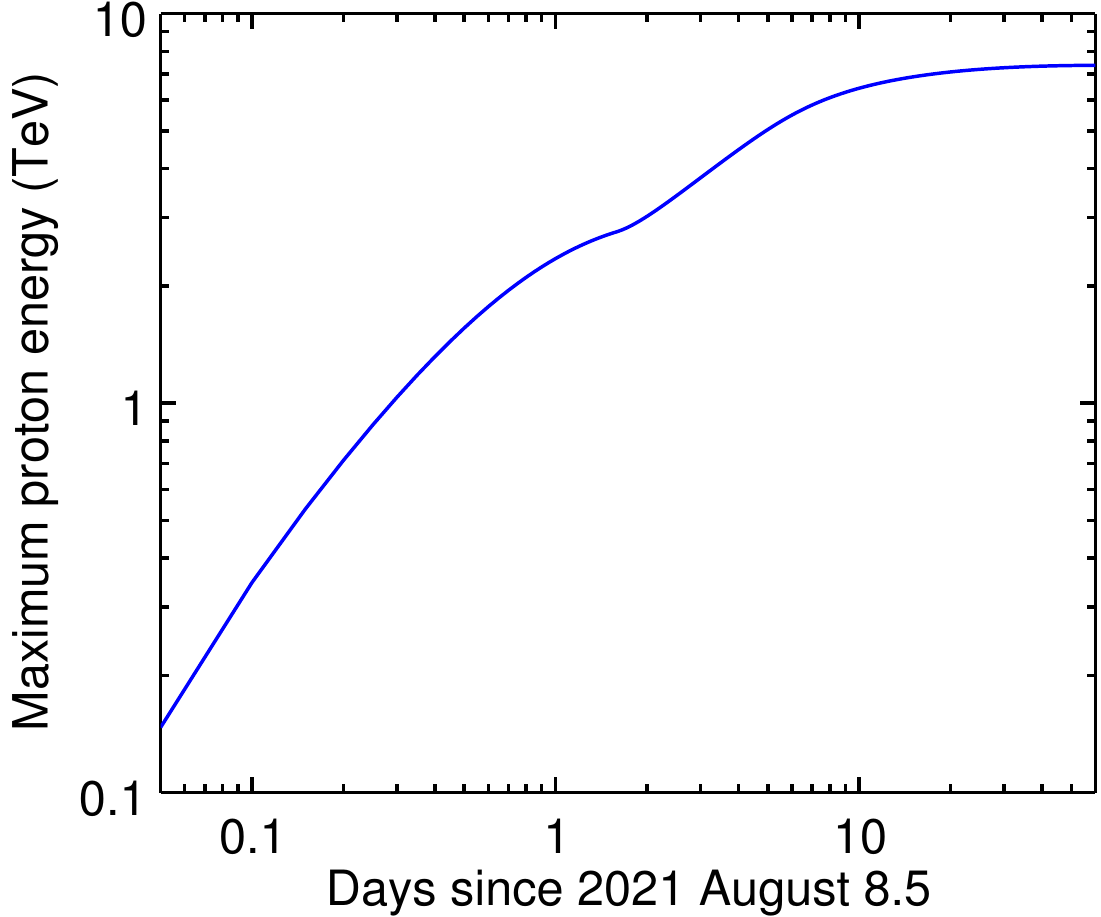}\hspace{0.1in}\includegraphics[scale=0.79,angle=0]{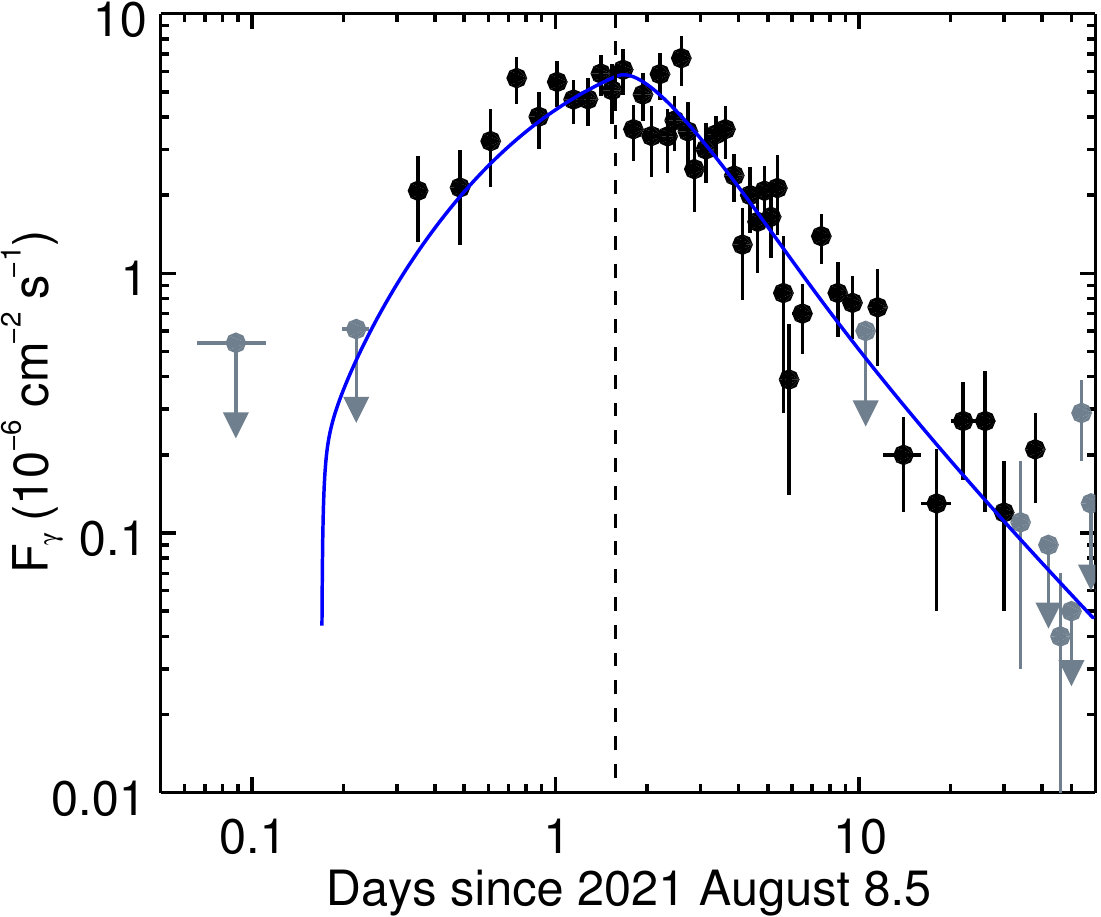}
\caption{(Left): Maximum proton energy with time for the hadronic model for RS Oph 2021.
(Right): Comparison of the hadronic model with the observed \gray\ flux light-curve presented in Figure~\ref{fig:gammaopticalxray}.
The dashed line shows the date at which the ejecta reach the radius of 1.5$\times$ the binary separation (see text).}
\label{fig:modrsophlc}
\end{center}
\end{figure}

In the context of the hadronic model, we define four emission phases for spectral study with the $>$50\,MeV LAT data (Figure~\ref{fig:pi0spectra}) -- the Rise from day 0 to 1.0, a Peak from day 1.0 to 2.75, Decline-a from day 3.0 to 9.0 (2--10$\times$ smaller than the peak), and Decline-b from day 9.0 to 46.0. 
The aim is to compare the evolution of the flux and spectral properties of the \gray\ emission in defined observation periods that are long enough to have sufficient statistics to determine accurate values of the spectral parameters. 
The Rise phase in particular was extended to include the initial portion of the peak to increase statistics in that bin (cf., Figure~\ref{fig:modrsophlc}, right).

The \gray\ spectrum of the hadronic model is calculated following the method described in \citet{kam06}, assuming that the energy distribution of the high-energy protons is a power law in proton momentum multiplied by an exponential cutoff \citep[see the supplementary material of][]{ack14}. 
The parameters fit to the LAT data are the normalization, the slope ($s_{\rm p}$) and the cutoff energy ($E_{\rm cp}$) of the high-energy proton spectrum. 
The fits for the four intervals were performed by maximizing the likelihood and the best-fit normalization is used to calculate the \gray\ spectral energy distributions (SEDs; Figure~\ref{fig:pi0spectra}). 
The corresponding best-fit parameters obtained for each observation period are presented in Table~\ref{table:spectra4intervals} along with the difference between the TS values resulting from the best-fit hadronic and ECPL models ($\Delta$TS).
According to the $\Delta$TS values, the hadronic model provides a fit as good as the one obtained with the ECPL model for most of the considered observation periods except for the Decline-a period where the former is preferred over the latter.
Between these four observation periods, the best-fit ECPL spectral parameters do not change significantly within the uncertainties which suggests that there is no significant variation in the spectral shape. 
Comparing the individual values to the averaged best-fit ones from the first 10-days after the outburst ($\Gamma\,=\,1.66\,\pm\,0.05$, $E_{\rm cut}\,=\,6.0\,\pm\,1.2$\,GeV; \S~\ref{sec:lat}), suggests a modest spectral change in the fit for the subset of data for the peak period ($E_{\rm cut}\,=\,3.16^{+1.10}_{-0.07}$\,GeV).
The slopes of the proton spectra are compatible with a constant value of $\sim$2.4 and there is no significant energy cutoff (i.e., lower limits are derived).

\begin{table*}
\begin{center}
\caption{\Fermi-LAT \gray\ spectral and hadronic model fit results for RS Oph 2021}
\hspace{-0.5in}
\begin{tabular}{lcccc}
\hline
\hline
Phase & Rise & Peak & Decline-a & Decline-b \\
 Days & 0.0 to 1.0 & 1.0 to 2.75 & 3.0 to 9.0 & 9.0 to 46.0 \\
\hline
\hline
\multicolumn{5}{c}{Exponentially Cutoff Power Law} \\
\hline
Photon Index, $\Gamma$	& $1.89 \pm 0.15$		& $1.50^{+0.09}_{-0.14}$ 	& $1.89 \pm 0.05$		& $1.48 \pm 0.34$	\\
$E_{\rm cut}$ (GeV)		& $11.4 \pm 12.1$  		& $3.16^{+1.10}_{-0.07}$	& $78.6 \pm 62.8$		& $3.39 \pm 2.69$ 	\\
Flux\,(0.05--300\,GeV) 	& $3.66 \pm 0.64$ 		& $7.36^{+0.37}_{-0.29}$ 	& $2.45 \pm 0.20$ 		& $0.17 \pm 0.06$  	\\
TS					& 212.2				& 1585.1				& 1012.3 				& 135.2			\\
\hline
\hline
\multicolumn{5}{c}{Hadronic Model} \\
\hline
Slope, $s_{\rm p}$ 		& $2.70^{+0.16}_{-0.27}$ 	& $2.30^{+0.23}_{-0.37}$ 	& $2.35^{+0.19}_{-0.23}$ 	& $2.15^{+0.32}_{-1.14}$ \\
$E_{\rm cp}\,{\rm(GeV)}$ 	& $>$46 				& $>$28 				& $>$63 				& $>$8 \\ 
Flux\,(0.05--300\,GeV) 	& $2.56^{+0.18}_{-0.23}$ 	& $6.02^{+0.26}_{-0.33}$ 	& $1.60^{+0.09}_{-0.08}$ 	& $0.15^{+0.02}_{-0.01}$ \\
${\rm \Delta TS}$ 		& --2.3 				& 8.9 				& 38.0 				& --3.5 \\
\hline
\end{tabular} 
\label{table:spectra4intervals} 
\end{center}
{\bf Notes.} The observation phases are defined in \S~\ref{sec:piondecay}. 
For the ECPL model fit results, $\Gamma$ and $E_{\rm cut}$ are the best-fit photon index and the cutoff energy of the photon spectrum, respectively.
For the hadronic model, the best-fit slope, $s_{\rm p}$ and cutoff energy, $E_{\rm cp}$ (2$\sigma$ lower limits) of the high-energy proton spectrum are given.
The fluxes at $>$50\,MeV energies are in units of \stphflux.
Values are given with their 1$\sigma$ statistical uncertainties. 
$\Delta$TS is the difference between the TS values of the hadronic model and the ECPL model.
\end{table*}

The hadronic model can explain the measured LAT spectra of RS Oph 2021 (see Figure~\ref{fig:pi0spectra}) and has also been proposed to explain its very high energy (VHE; $>$0.1\,TeV) \gray\ emission \citep{hes22,mag22}. 
Therefore, we constructed a simplified light-curve model assuming that the \gray\ emission results from \p0-decay produced by high-energy protons interacting with the material of the ejecta as proposed by \citet{tat07} to model the RS Oph 2006 outburst; see also \citet{her12}. 
The protons are accelerated via the Fermi process in the shock between the nova ejecta as it propagates through the RG wind. 
We estimated the speed of the ejecta with time using an analytical model fitted to the velocities derived from the X-ray temperatures \citep[see][]{bod06} measured with the \Swift-XRT data by \citet{pag22} -- see details in Appendix~\ref{sec:swift}.
This velocity model is representative of the average variation of the shock velocity with time and is used to compute the time evolution of the nova shell radius. 
The model can be roughly described by a constant value of 2470\,km\,s$^{-1}$ for $t\,<\,6$ days and proportional to $t^{-0.43}$ at $t\,>\,6$ day \citep[which is close to the $t^{-0.5}$ variation used by][]{tat07}.
The density profile of the RG wind is taken from \citet{tat07}. 
For simplicity, the RG wind density is assumed to be uniform within a radius of $<$1.5$\times$ the binary separation of 1.5.\,AU \citep{fek00} because of the complexity of the matter distribution inside and around the binary system. 

We calculated the maximum energy of accelerated protons as a function of time (Figure~\ref{fig:modrsophlc}, left) by integrating the sum of energy loss and gain rates. 
The energy gain rate is computed as in \citet{tat07}, with a compression ratio of the shock of 4 (strong adiabatic shock) and a magnetic field estimated assuming equipartition in the compressed gas (from the RG wind) with a wind temperature of 10$^{4}$\,K. 
The energy loss rate takes into account Coulomb collisions and inelastic $p$-$p$ collisions in the compressed gas. 
They were estimated from the methods described in \citet{man94} and \citet{der13}, respectively. 
In our model, the maximum proton energy as a function of time is similar to that in \citet[][figure~2, therein]{tat07} except the maximum energy is $\sim$7.4\,TeV instead of $\sim$3\,TeV because in our case the acceleration started before day 1, while it started at day 1 in \citet{tat07}.
The maximum proton energies estimated are $>$1\,TeV starting at 0.3 days after the outburst, in agreement with the lower limits on the cutoff energy of the proton spectrum derived from the LAT data (see Table~\ref{table:spectra4intervals}). 
In our model, the maximum proton energy changes from $\sim$2 to $\sim$5\,TeV between days 1.3 to 5.4, the times spanned by the early H.E.S.S.~VHE observation periods of RS Oph 2021. 
This could explain the hardening observed in the VHE spectra and the increasing high-energy photon detections from $\sim$0.3 to 1 TeV between these two epochs \citep{hes22}.

To calculate \gray\ light-curves for RS Oph 2021 in our model, we assumed that an injection fraction $f_{\rm inj}$ of the RG wind protons crossed by the shock is accelerated toward the expanding ejecta with a Fermi-type spectrum with a slope of --2.1, up to the maximum energy. 
This spectrum is used as a source function of a diffusion equation that estimates the time evolution of the proton spectrum, taking into account losses via inelastic collisions in the expanding ejecta.
We neglect escape losses which would have the effect of softening the proton spectrum as the highest-energy protons would be the first to escape the shock, and would also require a larger injection fraction to fit the measured \gray\ light curve.
The resulting spectra are then used to compute the rate of \p0\ production \citep[with the energy-dependent cross-section of][]{der86} produced in collisions with the hadrons in the expanding ejecta, which is assumed to have a uniform density. 
The \gray\ fluxes were computed for a distance of 1.6\,kpc and an ejecta mass of $\sim$1.1\,$\times$\,10$^{-6}$ M$_{\odot}$ \citep[see][]{bod06}. 
The resulting \gray\ model fluxes compare well with the observed 2021 LAT light-curve data (Figure~\ref{fig:modrsophlc}, right), and is similar to that obtained by \citet{her12} for the 2006 outburst.
Our model was obtained for an injection fraction $f_{\rm inj}\,\sim\,4\,\times\,10^{-6}$, lower than the value of \citet{tat07}. 
Considering the ejecta mass and velocity model adopted in our calculations, we estimated the attenuation of $\gamma$ rays due to Compton scattering and pair creation from interactions with nuclei in the expanding ejecta during the earliest stages when the gas densities are highest\footnote{The Compton scattering and pair production processes dominate for $\simlt$100\,MeV and $\simgt$100\,MeV photons, respectively.}  \citep[][section~3.2 therein]{mar18}.
According to our model, the attenuation amounts to a factor of $\sim$2 smaller observed flux at $\sim$400 MeV (the peak of the SED) at day 0.2 and is negligible after day 0.6, so does not fully explain the factor of $\sim$5 smaller \gray\ to optical luminosity ratio observed in the first day (see Figure~\ref{fig:gammaoptical}).
The smaller ratio could be more fully accounted for by the lower high-energy proton fluxes at the earliest acceleration phase in the shock development (see Figure~\ref{fig:modrsophlc}, left) resulting in a lower production of $\gamma$ rays.

\section{Discussion and Summary}
\label{sec:summary}

The anticipated \gray\ detection of the outburst of the recurrent nova RS Ophiuchi with \Fermi-LAT was realized in 2021 \citep{her12}. 
With a peak flux, $F_{\gamma}$\,($>$0.1\,GeV)\,$\simeq$\,6\,\stphflux, the RS Oph 2021 outburst is the brightest nova detected thus far in $\gamma$ rays by the LAT.
The previous brightest LAT-detected nova was the classical nova V906 Car 2018 with a peak $F_{\gamma}$\,($>$0.1\,GeV)\,$\simeq$\,4\,\stphflux\ \citep{jea18,ayd20}, although its early $\gamma$ rays were missed due to an anomaly with a \Fermi\ solar panel assembly motor \citep{abd22}.
The detection of $>$\,5\,GeV and up to $\sim$20--23\,GeV energy photons (\S~\ref{sec:he}) help constrain the high-energy portion of the SEDs in our modeling of select intervals.
RS Oph's brightness in the LAT band extended to higher energies, resulting in the first VHE detection of a nova \citep{hes22,mag22}.
The VHE observations constrain the maximum energy of the accelerated protons in this nova to be $\sim$10\,TeV \citep{hes22}, in agreement with the value derived in our semi-analytical model (see \S~\ref{sec:piondecay}).

The LAT-observed $>$0.1\,GeV $\gamma$ rays peaked at $\sim$1 day after the optical outburst and could reflect the time needed for the shock to accelerate enough protons to significant energy while it propagates through the inner part of the binary system. 
However, we cannot exclude that this timescale is also due to a change in the \gray\ attenuation produced by the material in which the shock propagates during the first day after optical eruption.
The LAT light-curve showed evidence, albeit at low significances, of factors of two fluctuations in the flux on timescales of 30 to 200 minutes during the first few days.
Such variability may be expected because of density variations of the swept-up material by the shock in the inner part of the binary system and/or changes in the outflow \citep[see][]{obr06,rup08,sok08,wal08,orl09}.
The decline of the \gray\ emission starting at days 2--3 could correspond to the timescale at which the shock enters a region where the density of the RG wind decreases with the shock radius.

The \gray\ detection of RS Oph can be compared to previous novae in symbiotic binaries detected by the LAT, particularly the first LAT-detected nova V407 Cyg 2010 \citep{abd10}.
The RS Oph LAT detection was expected based on external shocks from the nova ejecta evolving in the dense RG wind \citep{tat07,her12}, while the V407 Cyg detection was unanticipated because it was a less-studied symbiotic binary.
The 6-hr LAT peak flux, $F_{\gamma}$\,($>$0.1\,GeV)\,$\simeq$\,6\,\stphflux\ in RS Oph was $\sim$4$\times$ brighter than in V407 Cyg, consistent with the factor of $\sim$2 distance scaling if a 2.7\,kpc distance is adopted for the latter.
V407 Cyg was not detected in the VHE band despite past observations performed by Cherenkov telescopes \citep[see][]{ali12,ahn15,lop15}.
Other symbiotic recurrent novae with optical outbursts observed with the \Fermi-LAT were V745 Sco 2014 \citep[consisting of two low-significance 1-day detections coincident with its optical peak;][]{che14,fra18} and the $>$5$\sigma$ LAT detection of V3890 Sgr 2019 \citep{bus19}. 
Interestingly, the nova V1535 Sco 2015 was a low-significance ($2-3\sigma$) LAT detection as well \citep{fra18} and is proposed to be a symbiotic system \citep{sri15,lin17}. 
The latter three systems have distances at $\sim$6\,kpc and greater, so are broadly consistent with distance-scaled fluxes.

The majority of LAT detections are of classical novae, involving less-dense circumbinary environments from their main sequence companions, at a rate of $\sim$1\,year$^{-1}$ starting in 2012 \citep{ack14,che16}; see \citet{cho21} for a review. 
These observations demonstrated the importance of internal shocks inside the nova ejecta in producing $\gamma$ rays \citep[e.g.,][]{mar18,vur18}. 
That symbiotic recurrent novae ejecta must interact with the RG wind makes the case that both internal and external shocks are important in novae.
In RS Oph 2021, it appears that hadronic processes from the external shock discussed in \S~\ref{sec:piondecay} is the dominant component \citep[see][]{hes22,mag22}. 

With all the known symbiotic recurrent novae detected in outburst during the LAT era, the only known system remaining is T CrB, at a distance of only 0.8\,kpc. 
Its next outburst could be as soon as in 2023--2026 \citep{sch10,sch19,lun20}, and if the distance-scaled fluxes hold, should be remarkably bright with fluxes, $F_{\gamma}$\,($>$0.1\,GeV)\,$\simgt\,1\,\times\,10^{-5}$\,ph\,cm$^{-2}$\,s$^{-1}$ thus will be studied in remarkable detail with the LAT.
It is also conceivable that the LAT could detect the next RS Oph outburst because its recurrence interval has been as short as nine years (thus 2030 or later). 
In these cases \citep[see][for a discussion of RS Oph]{her08}, the radioactive decay emission at MeV energies could also be observed with the Compton Spectrometer and Imager \citep{tom22}, which will operate as an all-sky survey similar to \Fermi, and is expected to launch in 2026. 
This could build the most-complete MeV-GeV picture of the different \gray\ components of a nova evolution.

\begin{acknowledgments}
The \Fermi\ LAT Collaboration acknowledges generous ongoing support from a number of agencies and institutes that have supported both the development and the operation of the LAT as well as scientific data analysis.
These include the National Aeronautics and Space Administration and the Department of Energy in the United States, the Commissariat \`a l'Energie Atomique and the Centre National de la Recherche Scientifique / Institut National de Physique Nucl\'eaire et de Physique des Particules in France, the Agenzia Spaziale Italiana and the Istituto Nazionale di Fisica Nucleare in Italy, the Ministry of Education, Culture, Sports, Science and Technology (MEXT), High Energy Accelerator Research Organization (KEK) and Japan Aerospace Exploration Agency (JAXA) in Japan, and the K.~A.~Wallenberg Foundation, the Swedish Research Council and the Swedish National Space Board in Sweden.
 
Additional support for science analysis during the operations phase is gratefully acknowledged from the Istituto Nazionale di Astrofisica in Italy and the Centre National d'\'Etudes Spatiales in France. This work performed in part under DOE Contract DE-AC02-76SF00515.

Research at the Naval Research Laboratory is supported by NASA DPR S-15633-Y.
K.L.P.~and A.P.B.~acknowledge funding from the UK Space Agency.

This work made use of data supplied by the UK \Swift\ Science Data Centre at the University of Leicester.
We acknowledge with thanks the variable star observations from the AAVSO International Database contributed by observers worldwide and used in this research.

We thank the anonymous referee and A.~Azzollini for their comments on the manuscript.
\end{acknowledgments}

\facilities{Fermi, Swift, AAVSO}

\appendix

\section{Additional Details of the LAT Analysis}
\label{sec:appendixlat}

\subsection{Light-curve data and highest-energy LAT photon list}
\label{sec:appendixhe}

\begin{figure}[t]
\begin{center}
\includegraphics[scale=0.45,angle=0]{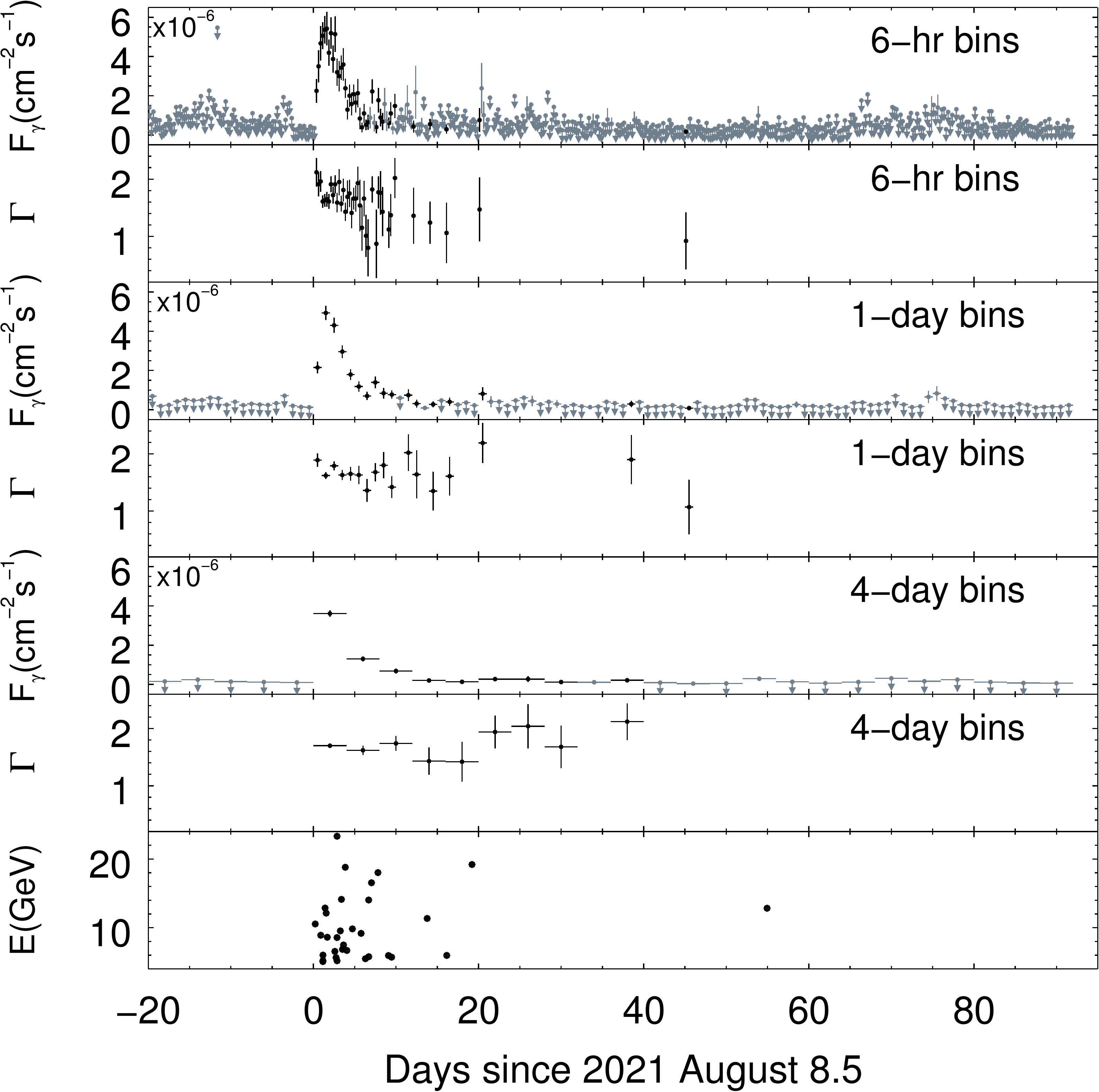}
\caption{From top to bottom: LAT $>$0.1\,GeV 6-hr light-curve and photon index, 1-day light-curve and photon index, 4-day light-curve and photon index, and energies of detected $>$5\,GeV photons.
Fluxes with error bars (black) are shown for TS\,$\geq$\,12 (3$\sigma$) points, while those with TS from 6.0 up to 12.0 (2--3$\sigma$) and upper limits are shown in gray.
Photon indexes are shown only for TS\,$\geq$\,12 points.}
\label{fig:lightcurvesLATall}
\end{center}
\end{figure}

The full results of the LAT $>$0.1\,GeV light-curve spectral analysis in 6-hr, 1-day, and 4-day bins described in \S~\ref{sec:lightcurves} are presented in Figure~\ref{fig:lightcurvesLATall}.

The list of LAT photons with $E\,>\,$5\,GeV found in the analysis described in \S~\ref{sec:he} is given in Table~\ref{table:HE} and plotted in Figure~\ref{fig:lightcurvesLATall} (bottom panel).

\begin{table*}
  \begin{center}
\caption{Highest-energy LAT photons selected at $E\,>\,$5\,GeV.}
\begin{tabular}{lcrc}
\hline
\hline
Day  & MET & \multicolumn{1}{c}{$E$} & gtsrcprob \\
        & (sec) & \multicolumn{1}{c}{(GeV)} & \\
\hline
 0.222	 & 650136003.9	 & 10.54 & 0.957 \\ 
 0.887	 & 650193475.7	 &  8.90 & 0.998 \\ 
 1.146	 & 650215820.9	 &  5.07 & 0.999 \\ 
 1.151	 & 650216236.4	 &  5.19 & 0.999 \\ 
 1.155	 & 650216600.8	 &  6.01 & 0.999 \\ 
 1.411	 & 650238679.9	 & 12.86 & 0.998 \\ 
 1.546	 & 650250356.7	 & 12.12 & 0.999 \\ 
 1.681	 & 650262005.6	 &  8.61 & 0.999 \\ 
 2.597	 & 650341199.0	 	&  6.55 & 0.968 \\ 
 2.740	 & 650353560.9	 &  5.65 & 1.000 \\ 
 2.861	 & 650363956.4	 & 23.32 & 0.999 \\ 
 2.861	 & 650363998.8	 &  8.57 & 0.997 \\ 
 2.863	 & 650364171.6	 &  5.18 & 0.969 \\ 
 3.260	 & 650398494.0	 &  9.53 & 0.999 \\ 
 3.391	 & 650409765.4	 & 14.12 & 0.999 \\ 
 3.518	 & 650420734.2	 &  6.84 & 0.999 \\ 
 3.645	 & 650431699.5	 &  7.48 & 0.996 \\ 
 3.859	 & 650450260.5	 & 18.80 & 0.971 \\ 
 4.049	 & 650466613.6	 &  6.69 & 0.992 \\ 
 4.721	 & 650524733.6	 &  9.83 & 1.000 \\ 
 5.774	 & 650615677.8	 &  9.17 & 0.996 \\ 
 6.295	 & 650660671.4	 &  5.48 & 0.999 \\ 
 6.692	 & 650694990.2	 & 14.03 & 1.000 \\ 
 6.703	 & 650695986.1	 &  5.78 & 0.998 \\ 
 7.034	 & 650724541.4	 & 16.54 & 0.881 \\ 
 7.815	 & 650791983.0	 & 18.02 & 0.916 \\ 
 9.090	 & 650902208.1	 &  5.95 & 0.999 \\ 
 9.471	 & 650935065.6	 &  5.70 & 0.999 \\ 
13.776	 & 651307020.6	 & 11.35 & 0.978 \\ 
16.147	 & 651511940.3	 	&  5.96 & 1.000 \\ 
19.205	 & 651776094.4	 & 19.20 & 0.996 \\ 
54.945	 & 654864060.0	 & 12.83 & 0.847 \\ 
\hline
\end{tabular}
\label{table:HE}
\end{center}
\smallskip
Note: Days are relative to \t0. Mission elapsed time (MET) for \Fermi\ is defined as seconds since 2001.0 UTC. 
\\
\end{table*}

\subsection{LAT Orbit-binned and Split-orbit spectral analysis}
\label{sec:appendixorbit}

\Fermi\ achieves its all-sky exposure profile by alternating between north and south scans of the sky during each spacecraft orbit ($\sim$1.6\,hr). 
The bulk of the exposure at the RS Oph position comes from the alternating southern sky scans.
For the LAT orbit-binned analysis, we define individual orbits by calculating the zenith angle of the center of our ROI, in 30-s steps, and looking for points where the angle reaches a maximum and then begins decreasing.  
We refer to these maxima as `orbit midnights' and use them to define orbit bin endpoints.  
We also calculate the angle ($\theta$) between the center of our ROI and the LAT z-axis and exclude from our final analysis those orbits where $\theta$ is always greater than 60\deg\ and those orbits with fewer than 10 events.  
In each orbit, we used the same radius ROI and energy selections described previously in Section~\ref{sec:lightcurves}.
When calculating the exposure, we considered the instrument azimuth angle, using \texttt{phibins=5} in the \texttt{fermitool} \texttt{gtltcube} and then performed an unbinned likelihood analysis, with only the isotropic diffuse emission component and the normalization and $\Gamma$ value of the nova free to vary.  
This approach resulted in 22 orbital bins (Table~\ref{table:orbit}), over the time period from \t0\ to \t0+3 days, with average exposure lengths of 57.5 minutes (0.04 days) and average offsets of the bin-centers of 190.4 minutes (0.13 days).  
The exposure lengths were estimated from an exposure light-curve for a 5\deg\ radius selection around our ROI center, in five minute bins, made using the \texttt{fermitools} \texttt{gtbin} and \texttt{gtexposure}.
All detections were TS\,$>$\,25 ($>$4$\sigma$ for 2 d.o.f.) with $95\%$ confidence upper limits derived for the first two bins (days 0.088 and 0.221).

We additionally conducted a ``split orbit'' analysis with \texttt{gtlike}, selecting nine orbit bins of particular interest.
The orbits include seven high-flux bins spanning the observed peak (days 0.745 and days 1.012 to 1.668), and two later high-flux bins (days 2.207 and 2.594). 
The exposure profile over an orbit is not generally uniform, so we divide it by examining the exposure accumulation and determining the time which yields equal exposure.
The results are presented in Table~\ref{table:orbitsplits}.
The most-significant variability was a factor of two drop at day 1.41 as described in Section~\ref{sec:shortest}. 
For a single orbit (days 0.745), the photon index $\Gamma$ appears to increase by $1.0\,\pm\,0.5$ ($\sim$2$\sigma$).

\begin{table*}
  \begin{center}
  \caption{Orbit-binned LAT analysis in the first three days.}
\begin{tabular}{lccccc}
\hline
\hline
Day   & $\Delta$T & TS & $N_{\rm pred}$ & Flux  (0.1--300\,GeV) & $\Gamma$ \\
         & (day) & & & (\stphflux) & \\
\hline
0.088	 & 0.023	 &   0 &  0.0 & $<$0.5	 & --	 \\ 
0.221	 & 0.023	 &  5.3 &  1.4 & $<$0.6	 & --	 \\ 
0.352	 & 0.023	 &  33 & 18.0 & $2.1 \pm 0.8$	 & $2.1 \pm 0.3$	 \\ 
0.484	 & 0.023	 &  28 & 17.5 & $2.1 \pm 0.8$	 & $2.1 \pm 0.3$	 \\ 
0.611	 & 0.017	 &  36 & 20.2 & $3.2 \pm 1.1$	 & $2.0 \pm 0.3$	 \\ 
0.745	 & 0.019	 &  95 & 39.6 & $5.7 \pm 1.2$	 & $2.3 \pm 0.2$	 \\ 
0.880	 & 0.023	 &  92 & 38.0 & $4.0 \pm 1.0$	 & $1.7 \pm 0.2$	 \\ 
1.012	 & 0.023	 & 140 & 48.4 & $5.5 \pm 1.1$	 & $1.9 \pm 0.2$	 \\ 
1.146	 & 0.021	 & 230 & 49.1 & $4.7 \pm 0.9$	 & $1.4 \pm 0.1$	 \\ 
1.278	 & 0.021	 & 120 & 43.7 & $4.7 \pm 1.0$	 & $1.7 \pm 0.2$	 \\ 
1.410	 & 0.021	 & 180 & 54.9 & $5.9 \pm 1.0$	 & $1.6 \pm 0.1$	 \\ 
1.535	 & 0.014	 & 120 & 29.0 & $5.1 \pm 1.3$	 & $1.7 \pm 0.2$	 \\ 
1.668	 & 0.016	 & 150 & 42.5 & $6.1 \pm 1.2$	 & $1.7 \pm 0.2$	 \\ 
1.804	 & 0.019	 & 110 & 32.2 & $3.6 \pm 0.9$	 & $1.5 \pm 0.2$	 \\ 
1.939	 & 0.023	 & 130 & 43.1 & $4.9 \pm 1.0$	 & $1.8 \pm 0.2$	 \\ 
2.071	 & 0.023	 &  49 & 27.3 & $3.4 \pm 1.0$	 & $2.1 \pm 0.3$	 \\ 
2.207	 & 0.019	 & 150 & 44.9 & $5.8 \pm 1.2$	 & $1.9 \pm 0.2$	 \\ 
2.338	 & 0.019	 &  58 & 26.5 & $3.4 \pm 0.9$	 & $1.8 \pm 0.2$	 \\ 
2.467	 & 0.023	 &  82 & 32.2 & $3.9 \pm 0.9$	 & $1.7 \pm 0.2$	 \\ 
2.594	 & 0.014	 &  99 & 34.6 & $6.7 \pm 1.5$	 & $2.0 \pm 0.2$	 \\ 
2.727	 & 0.016	 &  62 & 23.7 & $3.5 \pm 1.0$	 & $1.9 \pm 0.2$	 \\ 
2.865	 & 0.021	 &  93 & 24.0 & $2.5 \pm 0.8$	 & $1.4 \pm 0.2$	 \\ 
\hline
\end{tabular}
\label{table:orbit}
\end{center}
\smallskip
{\bf Notes.} Days are relative to \t0, and $\Delta$T represents the half binwidth.
\end{table*}

\begin{table*}
  \begin{center}
  \caption{Split orbit-binned LAT analysis results for select orbits.}
\begin{tabular}{lcccc}
\hline
\hline
Day   & $\Delta$T & TS & Flux (0.1--300\,GeV) & $\Gamma$ \\
         & (day) & & (\stphflux) & \\
\hline
0.7378	 & 0.0121	 &  57 & $4.5 \pm 1.5$	 & $1.9 \pm 0.3$	 \\ 
0.7569	 & 0.0070	 &  42 & $6.4 \pm 1.8$	 & $2.9 \pm 0.5$	 \\ 
1.0026	 & 0.0130	 &  93 & $5.8 \pm 1.4$	 & $1.8 \pm 0.2$	 \\ 
1.0251	 & 0.0096	 &  45 & $5.6 \pm 1.6$	 & $2.3 \pm 0.3$	 \\ 
1.1362	 & 0.0113	 & 114 & $4.4 \pm 1.2$	 & $1.3 \pm 0.2$	 \\ 
1.1571	 & 0.0096	 & 119 & $5.0 \pm 1.3$	 & $1.5 \pm 0.2$	 \\ 
1.2683	 & 0.0113	 &  43 & $4.2 \pm 1.4$	 & $1.8 \pm 0.3$	 \\ 
1.2891	 & 0.0095	 &  80 & $5.6 \pm 1.4$	 & $1.8 \pm 0.2$	 \\ 
1.4003	 & 0.0114	 & 125 & $8.1 \pm 1.8$	 & $1.6 \pm 0.2$	 \\ 
1.4211	 & 0.0095	 &  60 & $4.0 \pm 1.2$	 & $1.5 \pm 0.2$	 \\ 
1.5304	 & 0.0095	 &  61 & $6.7 \pm 2.2$	 & $1.8 \pm 0.3$	 \\ 
1.5443	 & 0.0044	 &  62 & $4.2 \pm 1.6$	 & $1.5 \pm 0.3$	 \\ 
1.6632	 & 0.0104	 &  48 & $5.6 \pm 1.8$	 & $1.8 \pm 0.3$	 \\ 
1.6788	 & 0.0052	 & 103 & $6.5 \pm 1.6$	 & $1.6 \pm 0.2$	 \\ 
2.1970	 & 0.0095	 &  73 & $6.9 \pm 1.7$	 & $2.0 \pm 0.2$	 \\ 
2.2161	 & 0.0096	 &  75 & $5.8 \pm 1.8$	 & $1.8 \pm 0.3$	 \\ 
2.5894	 & 0.0096	 &  50 & $6.7 \pm 2.1$	 & $1.8 \pm 0.3$	 \\ 
2.6033	 & 0.0043	 &  51 & $7.1 \pm 2.1$	 & $2.3 \pm 0.4$ 	\\ 
\hline
\end{tabular}
\label{table:orbitsplits}
\end{center}
\smallskip
{\bf Notes.} Days are relative to \t0, and $\Delta$T represents the half binwidth.
\end{table*}

\section{Optical Data}
\label{sec:opt}

\subsection{Optical Monitoring Data}
\label{sec:optical}

We utilized Visual-based measurements and optical photometry collected by the AAVSO for comparison to the LAT \gray\ light-curves (Figure~\ref{fig:gammaopticalxray}).
The optical nova discoveries from days 0.41--0.43 (2021 August 8.91--8.93; \S~\ref{sec:intro}) were at Visual magnitudes of $\sim$4.9--5.0.
Two earlier Visual observations at $\sim$5.1--5.2 mag from day 0.37 were later reported by J.~Manzorro and A.~Kosa-Kiss, and subsequent monitoring helped to detail the overall rising trend in the optical flux up to the peak on day $\sim$1.1.

Pre-discovery photometry data were obtained serendipitously with commercial cameras and reported to the AAVSO.
The earliest observations showing evidence of brightening were from four DSLR images obtained by \citet[AAVSO Observer Code WBIA, including observers L.P.~Lou and D.Y.~Chen]{wan21} indicating the initial rapid rise in brightness from $V$ = 9.1 to 7.0\,mag over a $\sim$100-minute span from day 0.075 to 0.144; a limit of $V\,>\,9.1$ was obtained from an observation on day 0.033.
From a linear extrapolation of these observations back to the quiescent brightness of $V$\,=\,11.1\,mag, \citet{mun21} derived a time of eruption of 2021 August 8.50 ($\pm$0.01) which we adopt here as the reference epoch (\t0) in the LAT analysis and discussion.
For the three weeks before the eruption, the quiescent brightnesses in the AAVSO database typically ranged from 11.1--11.3 mag, so the variations have only a small effect on the eruption epoch constraint.

\begin{figure}[t]
\begin{center}
\includegraphics[scale=0.8,angle=0]{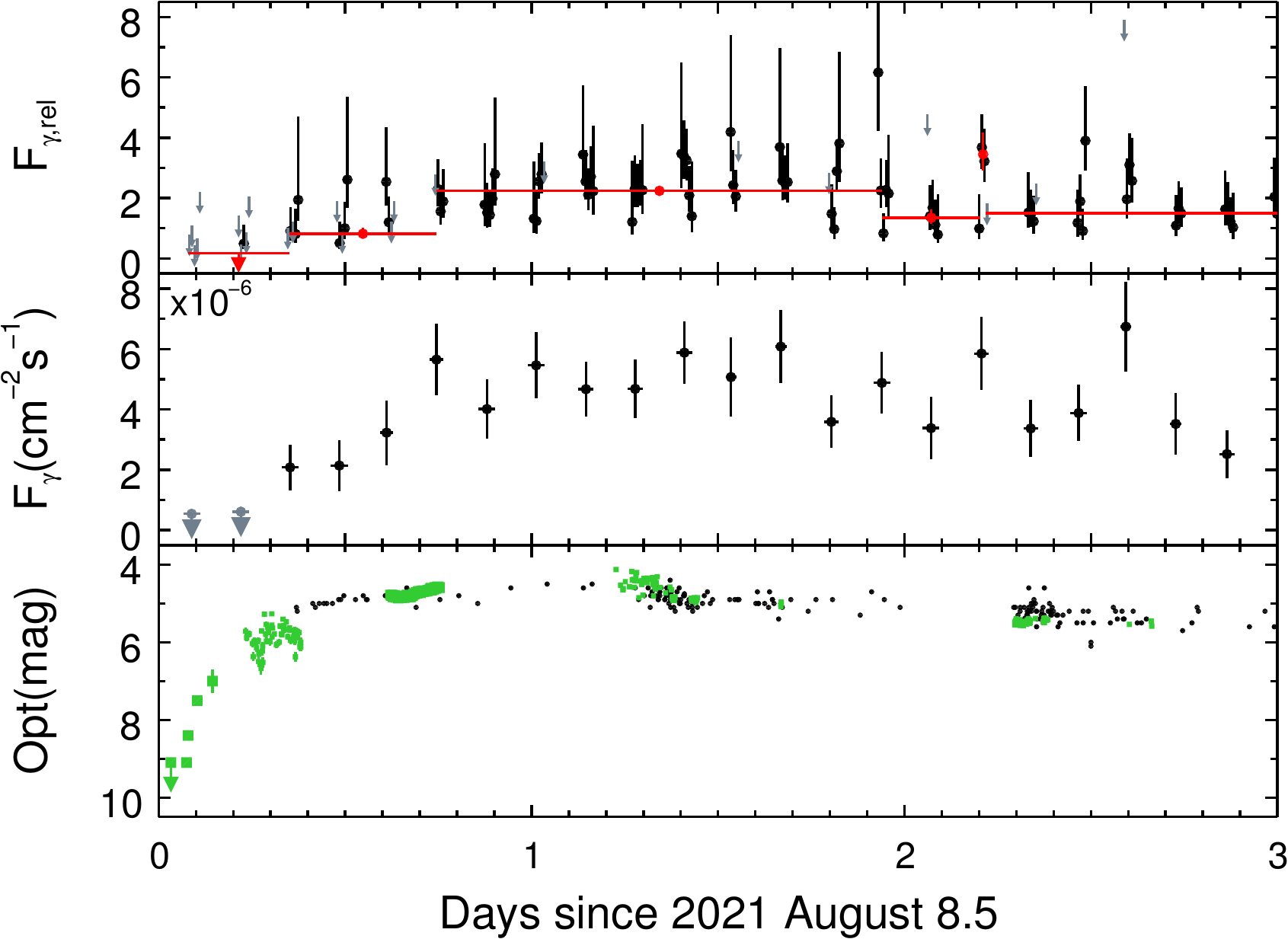}
\caption{Selected panels from Figure~\ref{fig:gammaplus} -- 
LAT $>$0.1\,GeV light-curve in 10-minute bins in units of relative flux and Bayesian Block partitions indicated in red, 
LAT $>$0.1\,GeV orbit-bin light-curve, and
optical light-curve, but for the first three days of activity. 
This version of the figure helps show more detail in the optical observations obtained from the Global Meteor Network camera IL0003 on the nights of August 8 and 9.}
\label{fig:gammaplusapp}
\end{center}
\end{figure}

The nova was also captured by the Global Meteor Network \citep{vid21} camera IL0003 located in Israel (by A.L.~and A.B.) with a 4\,mm f/0.95 lens and a Sony IMX291 sensor.
The 58 individual measurements from August 8, 17:35:46 to 21:09:05 (day 0.233 to 0.381) detailed further brightening\footnote{These are shown in place of the four averaged measurements reported by one of us (K.V.S.) to the AAVSO.} following the first four observations from \citet{wan21} up to the discovery observations.
The individual measurements for August 8 as well as further observations on August 9 (37 measurements from 17:26:19 to 21:00:29; day 1.227 to 1.375) are newly presented in this paper (see Figure~\ref{fig:gammaplusapp} in particular).
Aperture photometry with $V$ zeropoint on these unfiltered images (labelled $CV$ in the AAVSO) was performed using the VaST code \citep{sok18} assuming $V$\,=\,3.34 for $\nu$ Oph \citep{duc02}, and these are simply referred to as $V$-band measurements throughout this paper. 
The second night measurements overlapping in time with the Visual observations from the AAVSO are in good agreement considering the large scatter ($\sim$0.5 mag) in the meteor-camera observations due to the noisy and variable background structure visible across the CMOS images.

\subsection{Optical luminosities from observed magnitudes}
\label{sec:opticallum}

We also used available photometry collected by the AAVSO for RS Oph to calculate observed optical luminosities by assuming a Planck function spectral form with gray opacity to fit the photometric data at different epochs.
Magnitudes were corrected for extinction due to the interstellar medium (ISM) and we applied approximate blackbody temperature-dependent bolometric corrections \citep{wei67}.
The temperatures were estimated with two methods, similar to those used by \citet{li17} and \citet{ayd20} to analyze the time variation of the ratio of \gray\ to optical luminosities of classical novae.
Our aim was to use the fitted temperature as a parameter to estimate total optical energy fluxes, rather than deriving temperatures of the nova pseudo-photosphere.

\begin{figure}[t]
\begin{center}
\includegraphics[scale=0.7,angle=0]{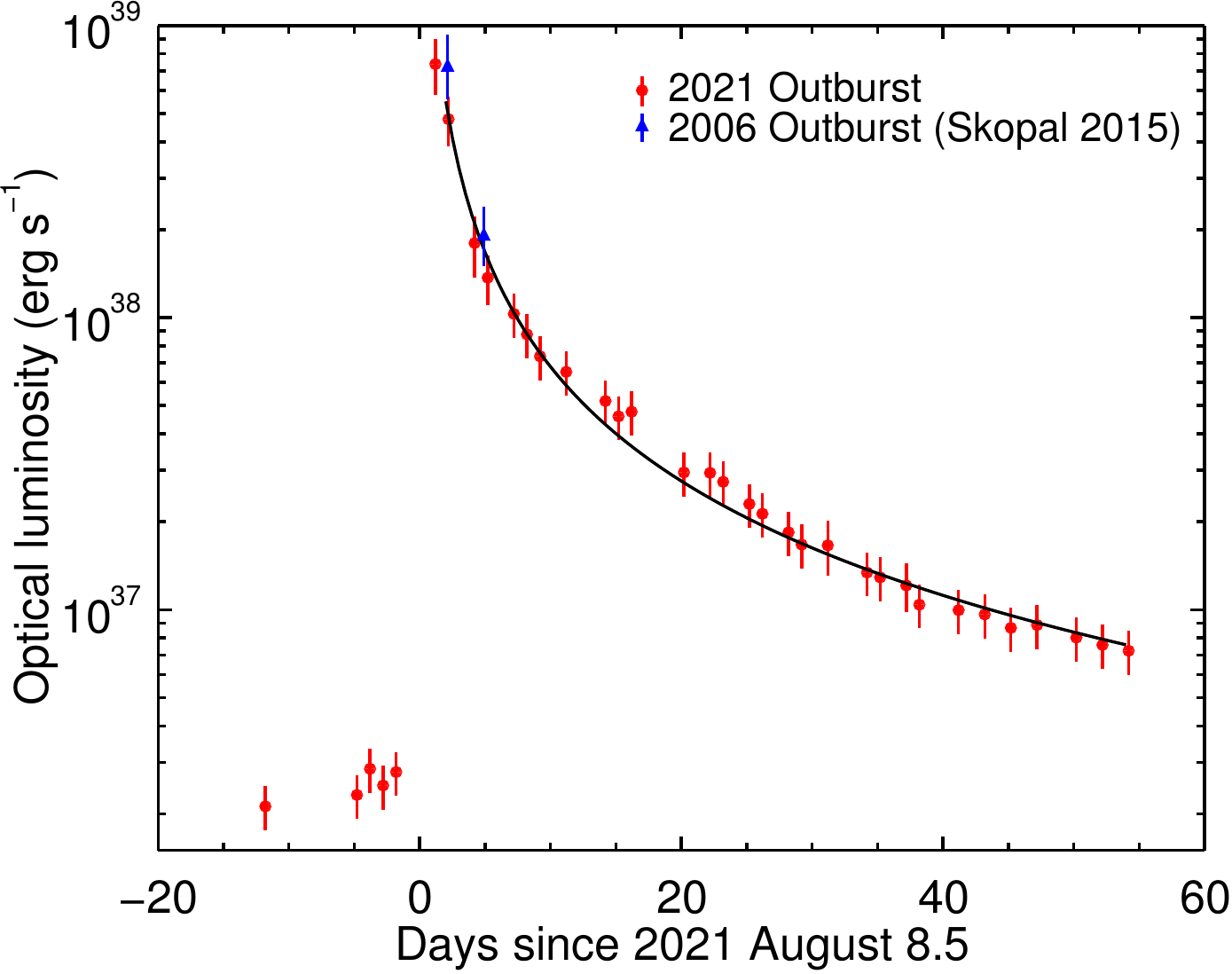}
\caption{Optical luminosities for the 2021 pre- and post-outburst emission from RS Oph derived from AAVSO data (red circles; see text).
The best-fit power law model for the decline (black line) has a slope of $-1.28\,\pm\,0.04$.
The optical luminosities estimated by \citet{sko15} for times near the peak of the previous 2006 outburst are shown for comparison (blue triangles; converted to our definition of \t0\ by adding 1.1 day).
}
\label{fig:opticallum}
\end{center}
\end{figure}

The first method involves fitting a functional blackbody temperature at several dates to the SED composed of observed $B$-, $V$-, $R$-, and $I$-band magnitudes available in the AAVSO\footnote{The approximate wavelengths for the filters are listed in: \url{https://www.aavso.org/filters}}, after correction of extinction.
The extinction correction for each band was calculated with a relative ISM extinction model from optical to mid-infrared bands obtained by \citet{wan19} with photometric data collected by several instruments.
This method requires quasi-simultaneous measurements in each of the spectral bands.
Such measurements were performed at only six dates from day 4 to day 34.
For all the observation periods, except the first one, the best-fit temperatures are $\approx$\,8000\,K, with uncertainties of $\approx$\,1000\,K. 
The quality of the SED fit in the first period (day 4) was poor thus the best-fit temperature was not used.
Instead, we adopted a temperature of 8500\,K, which is intermediate between the temperatures of $\sim$7000--10000\,K obtained by \citet{sko15} from modeling the multi-wavelength SEDs measured during the first days of the 2006 outburst of RS Oph (see Figure~\ref{fig:opticallum}).

The second method is to calculate color temperatures, $T_{\rm c}$, using quasi-simultaneous $V$- and $B$-band magnitude measurements and two-color indices corrected for extinction, as done by \citet{li17}.
Extinction was derived from the reddening $E(B-V)$\,=\,0.73\,$\pm$\,0.06 \citep{sni87} and the extinction ratio, $R_{V}$\,=\,3.1. 
The $V$- and $B$-band measurements were obtained at 35 epochs from days $\sim$--12 to $\sim$55 (from AAVSO observers WGR, BDG, and FJQ).
This method provides a range of $T_{\rm c}$ = 8000 to 14000\,K after the outburst date with an average of $T_{\rm c}\,\approx\,(11000\,\pm\,1700)$\,K. 
Before the outburst, the color temperature was $T_{\rm c}\,=\,(5900\,\pm\,400)$\,K.

Figure~\ref{fig:opticallum} shows the thus-obtained variation of the observed optical luminosity of nova RS Oph 2021 before and after its outburst obtained from daily-averaged $V$- and $B$-band measurements from the AAVSO database.
The values close to the luminosity peak are similar to the ones obtained by \citet{sko15} for the first days after the 2006 outburst.
The luminosity appears to vary as a power-law in time from day 2 onward with a best-fit slope of $-1.28\,\pm\,0.04$.

\section{\Swift\ X-ray Monitoring Data}
\label{sec:swift}

\Swift\ began observing RS Oph on 2021 August 9 \citep{pag21c}, continuing until the season ended on November 4, after which RS Oph became Sun constrained (observable again on 2022 February 1). 
Early observations were obtained on a daily basis, the cadence was increased to twice a day from the start of September, and then to every 8\,hr between September 15 and October 1. 
Daily cadence then resumed until the observations ended. 
In addition, a very high cadence observing campaign (every \Swift\ 1.5-hr orbit) was performed on September 12 to investigate potential supersoft source variability. 
All of these observations were taken using Windowed Timing (WT) mode, since the XRT count rate always exceeded 1\,count s$^{-1}$. 
Despite the high count rate, however, there were a number of separate observations between September 24 and October 8 performed in Photon Counting (PC) mode specifically to investigate the point spread function of the source emission; only the WT data are discussed in this paper.

The data were analyzed using HEASoft 6.29 and the latest calibration available in 2021 November. 
The light-curve was produced using the online XRT product generator\footnote{\url{https://www.swift.ac.uk/user_objects/}} \citep{eva07,eva09}, using only grade 0 (single pixel) events to help minimize optical loading\footnote{See: \url{https://www.swift.ac.uk/analysis/xrt/optical_loading.php}}.
Relevant to the comparison to the \gray\ emission, we used the higher-energy rates at 2--10\,keV energies (Figures~\ref{fig:gammaopticalxray}~\&~\ref{fig:gammaplus}). 
In total, we found 121 significant detections in this band spanning days 1.4 to 87.6 from typical exposure times of 0.5--1\,ks.
The full XRT calibration and dataset, including the detailed discussion of the supersoft emission around day 27 \citep[using the \t0\ defined here;][]{pag21a,pag21b}, is presented elsewhere \citep{pag22}.

\begin{figure}[t]
\begin{center}
\includegraphics[scale=0.8,angle=0]{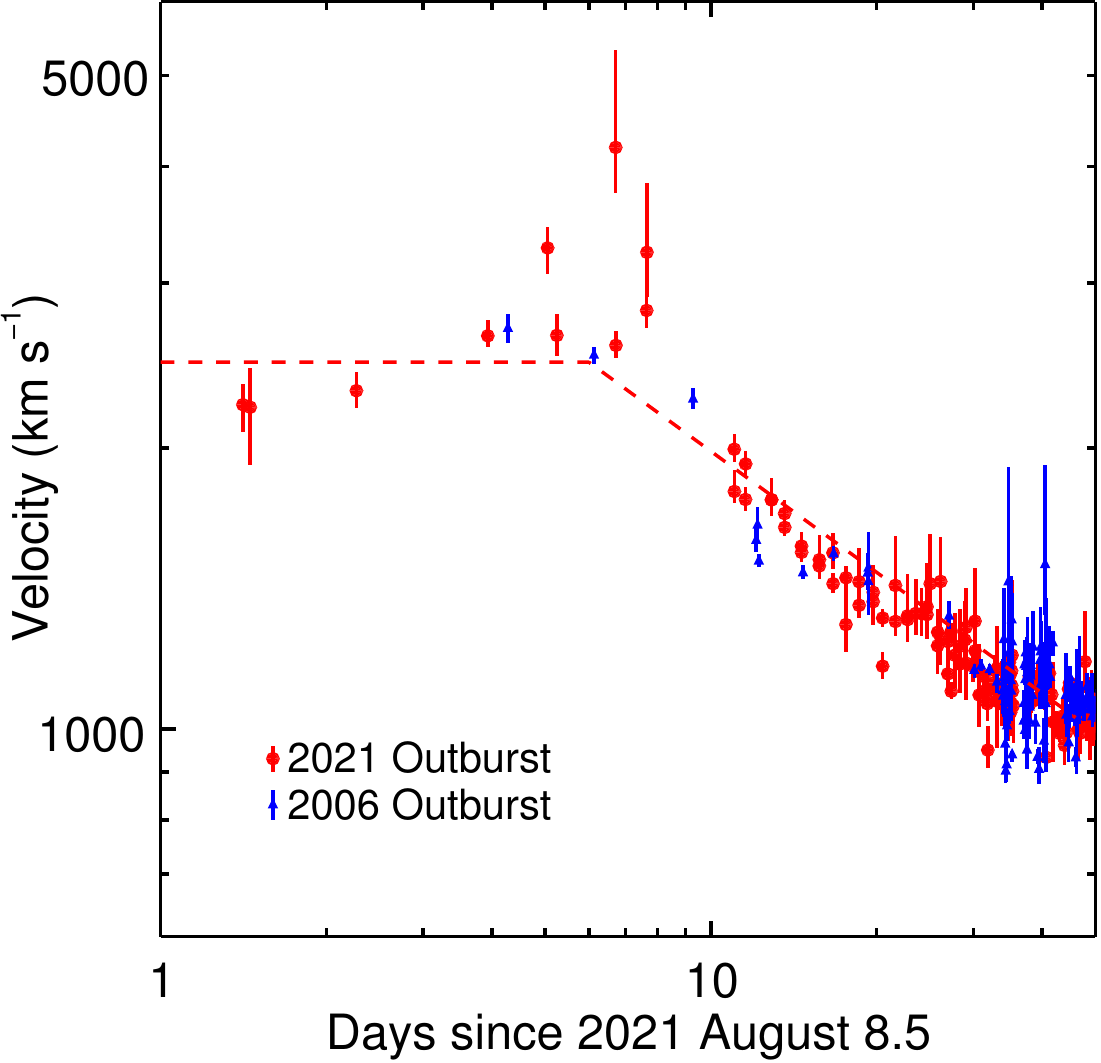}
\caption{Shock velocities for the nova ejecta from the 2021 outburst of RS Oph derived from \Swift-XRT temperature measurements (red circles; see text).
The dashed line corresponds to the profile of the 2021 outburst we adopted in our modeling, consisting of a constant value of 2470\,km\,s$^{-1}$ for $t\,<\,6$ days and proportional to $t^{-0.43}$ for $t\,>\,6$ day (see \S~\ref{sec:piondecay}). 
The velocities derived for the previous 2006 outburst are shown for comparison (blue triangles; converted to our definition of \t0\ by adding 1.1 day).
}
\label{fig:vel}
\end{center}
\end{figure}

In our modeling of the 2021 outburst presented in \S~\ref{sec:piondecay}, we used temperatures measured from spectral analysis of \Swift-XRT data to estimate the time-variation of the shock velocity of the nova ejecta \citep[see equation\,3 of][]{bod06}. 
The results of the spectral fits of the $0.3-10$\,keV XRT data were taken from \citet{pag22}, who adopted a model using two optically-thin plasma components to represent the shock emission, with an additional blackbody component at the lowest energies starting at day~$\sim$20.
To estimate the velocity, we used the temperature of the hotter of the two shock-related components \citep[$kT_{\rm hot}$ in][]{pag22}, which dominates the emission at $>$2\,keV energies (see figures 9 and 11, therein). 
The resultant velocities shown in Figure~\ref{fig:vel} only take the statistical uncertainties of the fitted temperatures into consideration; we do not account for systematic uncertainties related to the fits.
\citet{pag22} also reanalyzed the \Swift-XRT data for the 2006 outburst in the same way and we used those fitted $kT_{\rm hot}$ values to derive velocity data (only values of $kT_{\rm hot}$ constrained to a factor of 2 were used).
The obtained velocity profiles for the two explosions based on the \Swift-XRT data are similar (Figure~\ref{fig:vel}). 
Moreover, the XRT-derived velocity profile for the 2006 data is similar to the one used by \citet[][figure~1 therein]{tat07} to model that explosion based on the early \Swift-XRT results of \citet{bod06} and the \textit{Rossi} X-Ray Timing Explorer (RXTE) results from \citet{sok06}.

\end{document}